\documentclass[11pt]{article}
\usepackage{jcappub} 
\usepackage{color}
\usepackage{latexsym}
\usepackage{graphicx}
\usepackage{subfigure}
\usepackage{hyperref}
\usepackage{amssymb}
\usepackage{bm}
\usepackage{natbib}
\usepackage{amsmath}
\title{Cosmological signatures of tilted isocurvature perturbations: reionization and 21cm fluctuations}

\author[a,b]{Toyokazu Sekiguchi,}
\author[c]{Hiroyuki Tashiro,}
\author[d,e,f]{Joseph Silk,}
\author[a,g,h]{and Naoshi Sugiyama}

\affiliation[a]{Department of Physics and Astrophysics, Nagoya
University, Nagoya 464-8602, Japan}
\affiliation[b]{University of Helsinki and Helsinki Institute of Physics, 
P.O. Box 64, FI-00014, Helsinki, Finland}
\affiliation[c]{Physics Department, Arizona State University, Tempe AZ 85287, USA}

\affiliation[d]{Institut d'Astrophysique, UMR 7095 CNRS,\\
Universit\'{e} Pierre et Marie Curie,98bis Blvd Arago, 75014 Paris, France}
\affiliation[e]{Department of Physics and Astronomy, \\The Johns Hopkins University, Homewood Campus, Baltimore MD 21218, USA}
\affiliation[f]{Beecroft Institute of Particle Astrophysics and Cosmology, Department of Physics, \\University of Oxford, Oxford OX1 3RH, UK}

\affiliation[g]{Kobayashi-Maskawa Institute, Nagoya University, Nagoya 464-8602, Japan}
\affiliation[h]{
Kavli Institute for the Physics and Mathematics of the Universe (WPI), Todai Institutes for Advanced Study, The University of Tokyo,
Kashiwa 277-8568, Japan}

\emailAdd{toyokazu.sekiguchi@helsinki.fi}

\abstract{
We investigate cosmological signatures of uncorrelated
isocurvature perturbations whose power spectrum is blue-tilted with
spectral index $2\lesssim p\lesssim4$.  Such an isocurvature power
spectrum can promote early formation of small-scale structure, notably
dark matter halos and galaxies, and may thereby resolve the shortage of
ionizing photons suggested by observations of galaxies at high redshifts
($z\simeq 7-8$) but that are required to reionize the universe at $z\sim
10.$ We mainly focus on how the formation of dark matter halos can be
modified, and explore the connection between the spectral shape of CMB
anisotropies and the reionization optical depth as a powerful probe of a
highly blue-tilted isocurvature primordial power spectrum.  We also
study the consequences for 21cm line fluctuations due to neutral
hydrogens in minihalos.  Combination of measurements of the reionization
optical depth and 21cm line fluctuations will provide complementary
probes of a highly blue-tilted isocurvature power spectrum.  }

\begin{document}

\begin{flushright}
HIP-2013-26/TH
\end{flushright}

\maketitle

\section{Introduction}

The epoch of reionization is a milestone in cosmological structure
formation history.
Although recent cosmological observations are beginning to reveal the epoch of reionization,
the process of cosmological reionization is still one of the most open questions in cosmology.
In particular, observations of star-forming galaxies at high redshifts $6\lesssim z\lesssim 8$, 
suggest that the density of UV photons emitted from such galaxies may not be sufficient to keep the 
universe ionized~\cite{Robertson:2013bq}, if the number density of the galaxies are as predicted by 
the concordance cosmological model with a nearly scale-invariant curvature power spectrum 
consistent with current CMB observations~\cite{Hinshaw:2012aka,Ade:2013zuv},
unless very optimistic assumptions are made about ionizing photon escape fractions and other parameters. 
Indeed this conclusion is further aggravated by the decline in star formation rate density
recently found beyond $z\sim 8$~\cite{Oesch:2013bha}.

To enhance the formation of collapsed objects in the high redshift universe, 
one possible solution is to allow matter fluctuations to have a large amplitude at small scales.
However, in the inflationary cosmology, there are few degrees of freedom  available to 
generate large curvature perturbations at small scales since they are found to  be nearly-scale invariant
on the scales of CMB and galaxy clustering ($k\lesssim0.1$Mpc$^{-1}$). 
On the other hand, isocurvature perturbations can have a highly
blue-tilted power spectrum while satisfying CMB constraints on large 
scales\footnote{
For theoretical models, we refer to e.g. 
Refs.~\cite{Yokoyama:1990xv,Yokoyama:1995ex,Peebles:1998nu,Kasuya:2009up}.}.

In this study, we consider how such blue-tilted isocurvature power spectra can 
affect cosmological observables at high redshifts, and in particular the CMB via the reionization of the universe.
Such an isocurvature power spectrum
can promote the formation of dark matter halos and enhance the formation
of stars and galaxies in
the high-redshift universe.
While there are several previous studies on various aspects of these 
issues~\cite{Yokoyama:1990xv,Cen:1993uj,Peebles:1998ph,Sugiyama:2003tc}\footnote{
Blue-tilted isocurvature model can be also probed by CMB distortions~\cite{Dent:2012ne,Chluba:2013dna}.
},
we here focus on effects on reionization and 21cm fluctuations due to nonlinearities triggered by  isocurvature power spectra
and especially on the complementarity of these probes.
Although several authors have discussed effects of isocurvature power spectra on 21cm fluctuations~\cite{Gordon:2009wx, Kawasaki:2011ze}, 
they have generally focused on the contribution of a homogeneous IGM well before the epoch of reionization ($z\gtrsim20$).
Here we focus on collapsed halos which make dominant contributions
at relatively low redshifts ($z\lesssim20$).

Throughout this paper, we focus on matter isocurvature perturbations, where $S(\vec x)=\delta_m-\frac34\delta_\gamma$ ($\delta_i$ is the initial 
density perturbation in a component $i$) that is  uncorrelated with  curvature fluctuations, $\zeta(\vec x)$, 
i.e. $\langle S(\vec x)\zeta(\vec x')\rangle=0$. We in addition assume that there are no initial relative perturbations
between baryon and CDM or so-called {\it compensated isocurvature perturbations}~\cite{Gordon:2009wx}\footnote{
This is consistent with current data~\cite{Grin:2013uya}.}, i.e. $\delta_c=\delta_b$.
The power spectrum is defined as $\langle S(\vec k) S(\vec k')\rangle=P_{\rm iso}(k)(2\pi)^3\delta^{(3)}(\vec k+\vec k')$.
We assume the power spectrum $P_{\rm iso}(k)$ is in power-law form, so that it can be parameterized as 
\begin{equation}
P_{\rm iso}(k)=\frac{2\pi^2}{k^3}A_{\rm iso}\left(\frac{k}{k_0}\right)^{n_{\rm iso}-1}, 
\end{equation}
where $A_{\rm iso}$ and $n_{\rm iso}$ are the amplitude and the spectral index of $P_{\rm iso}(k)$ at the pivot scale $k_0=0.002$Mpc$^{-1}$, which
roughly corresponds to $\ell=10$ in the CMB angular power spectrum.
We define the ratio of the isocurvature to adiabatic power spectrum,
$\alpha=A_{\rm iso}/A_{\rm adi}$, where the adiabatic power
spectrum is given by
$P_{\rm adi}(k)=(2\pi^2/k^3)A_{\rm adi}(k/k_0)^{n_{\rm adi}-1}$.
Hereafter we 
use two parameters $\alpha$ and $p=n_{\rm iso}$ to characterize the model with 
an isocurvature power spectrum.

The CMB and matter power spectra are computed using the {\tt CAMB} code~\cite{Lewis:1999bs,Howlett:2012mh}.
We take the concordance flat $\Lambda$CDM model with power-law adiabatic power spectrum
as the baseline cosmological model and set  parameters to the best fit values from the WMAP 9-year 
result~\cite{Hinshaw:2012aka}.

This paper is organized as follows. To begin with, we briefly discuss current 
constraints on the blue-tilted isocurvature power spectrum from the shape of the CMB 
power spectrum in the next section. In order to see how such an isocurvature power spectrum
affects the late-time universe, we first investigate its effects on the
mass function of dark matter halos in Section~\ref{sec:massf}.
For observational signatures, we focus on the reionization optical depth and 21cm line fluctuations, which
are  discussed in Section~\ref{sec:opt} and \ref{sec:21cm}, respectively. 
A final section is devoted to our conclusions.

\section{CMB power spectrum and constraints }
\label{sec:CMB}

\begin{figure}
  \begin{center}
  \begin{tabular}{ccc}
    \hspace{-5mm}\scalebox{.4}{\includegraphics{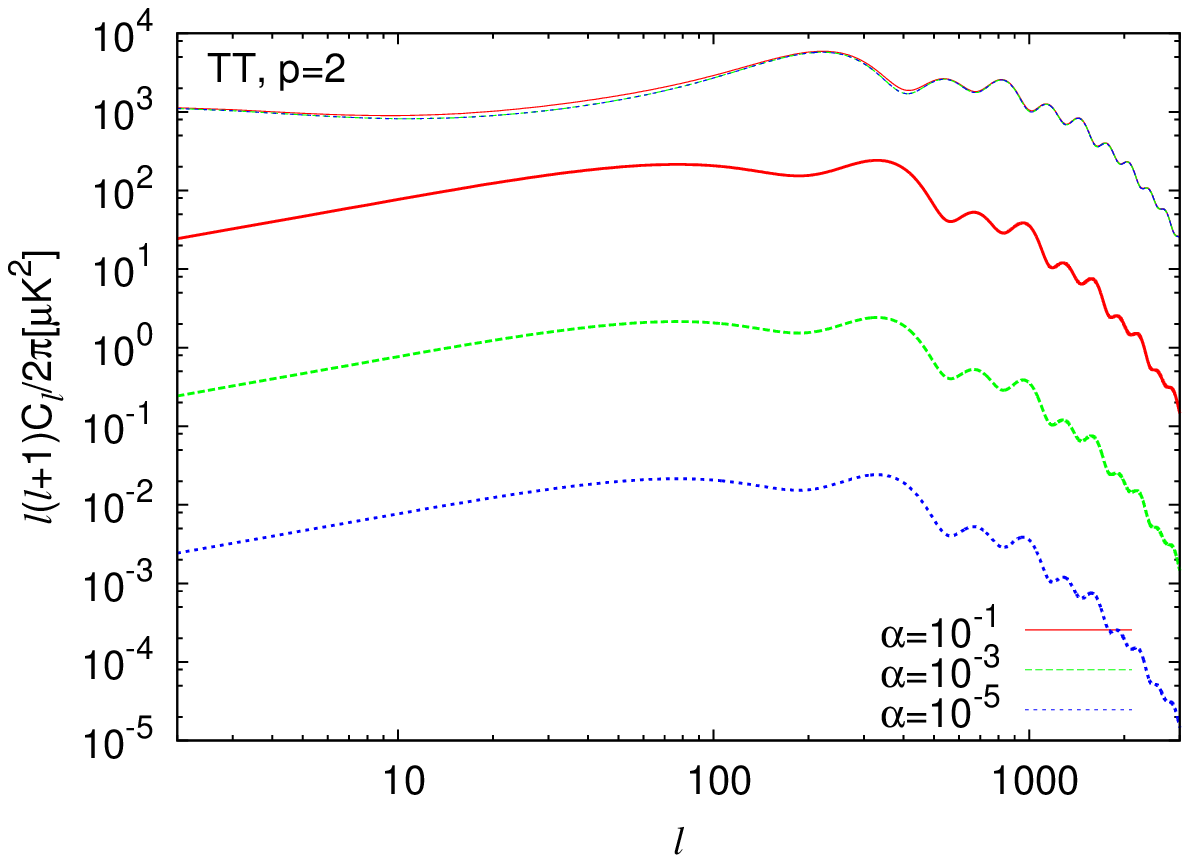}} &
    \hspace{-5mm}\scalebox{.4}{\includegraphics{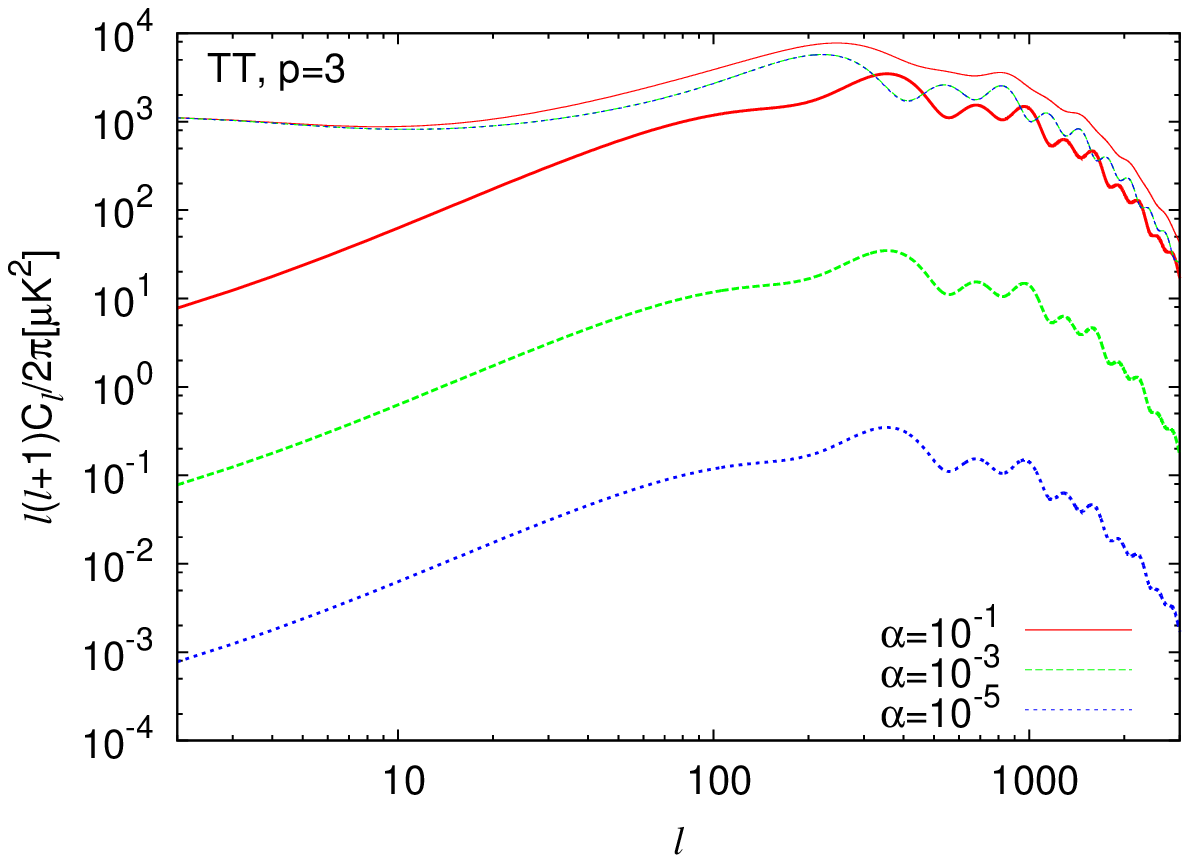}} &
    \hspace{-5mm}\scalebox{.4}{\includegraphics{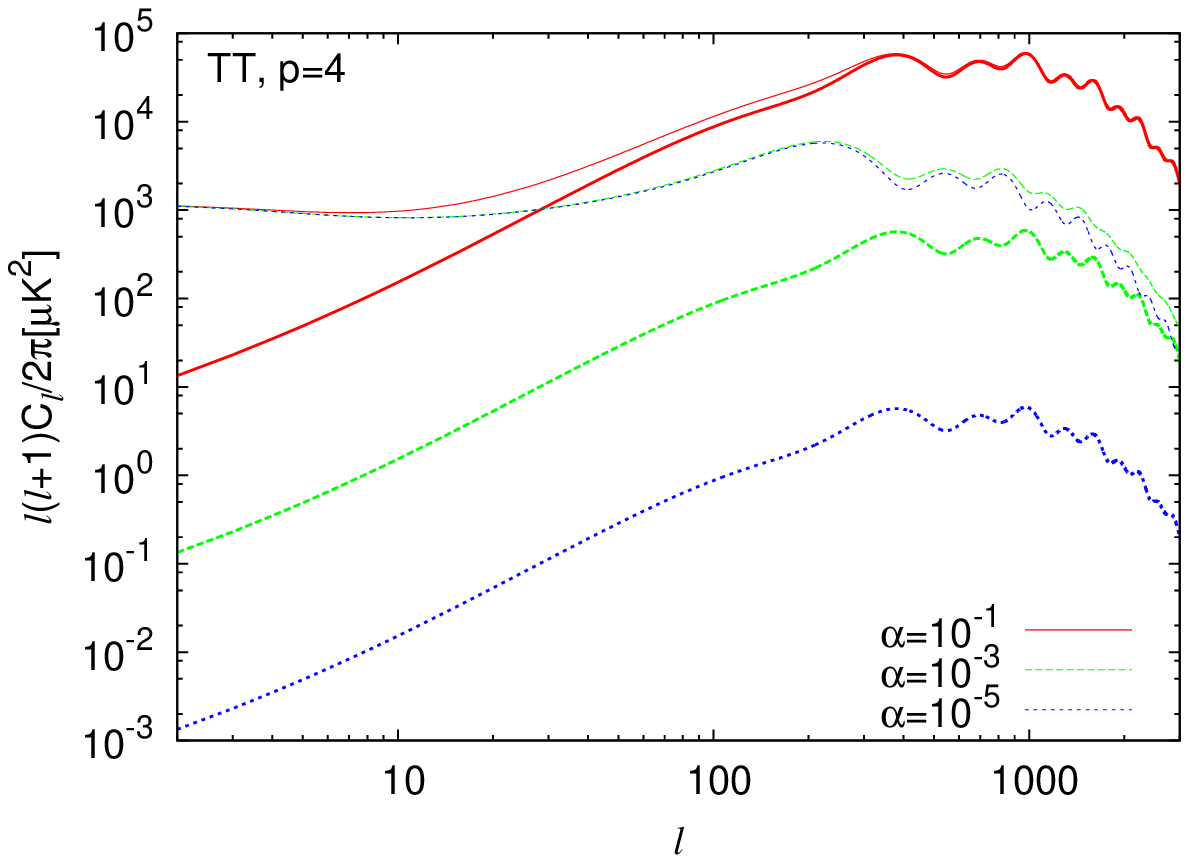}} \\
    \hspace{-5mm}\scalebox{.4}{\includegraphics{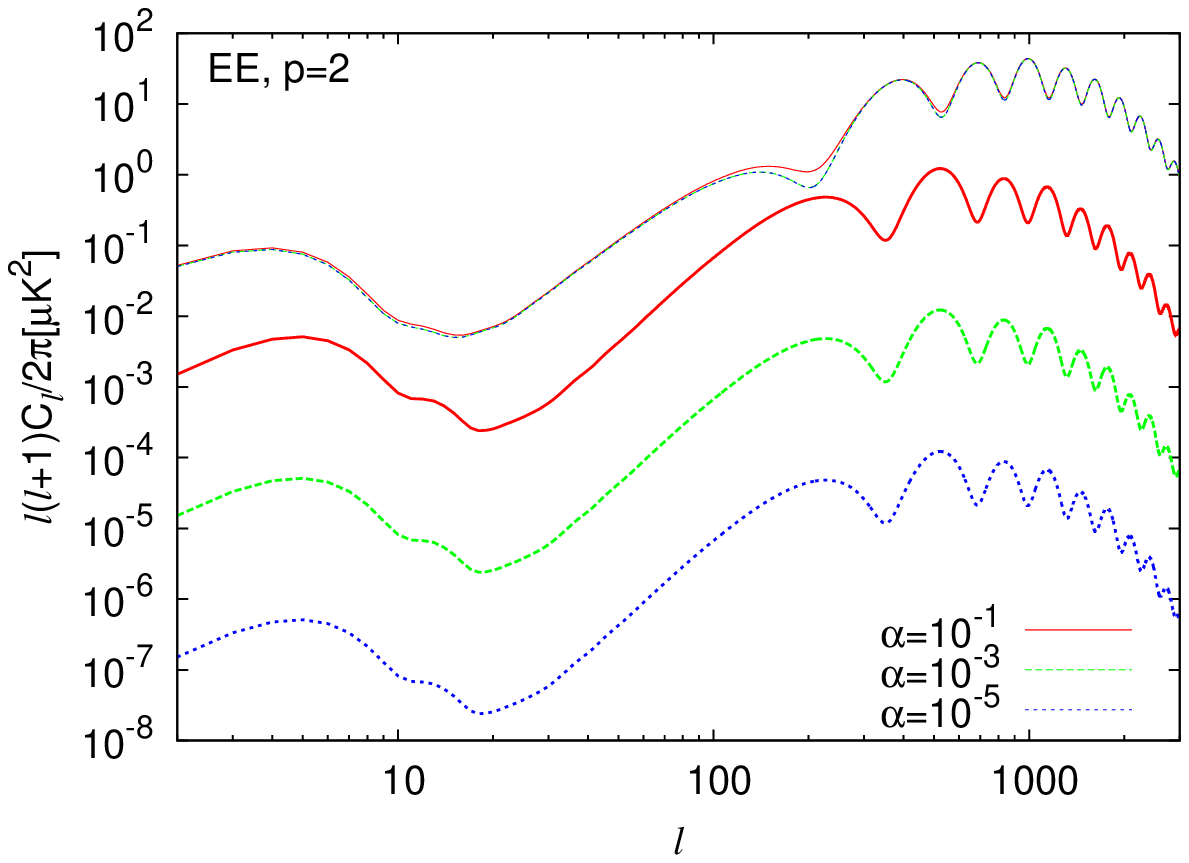}} &
    \hspace{-5mm}\scalebox{.4}{\includegraphics{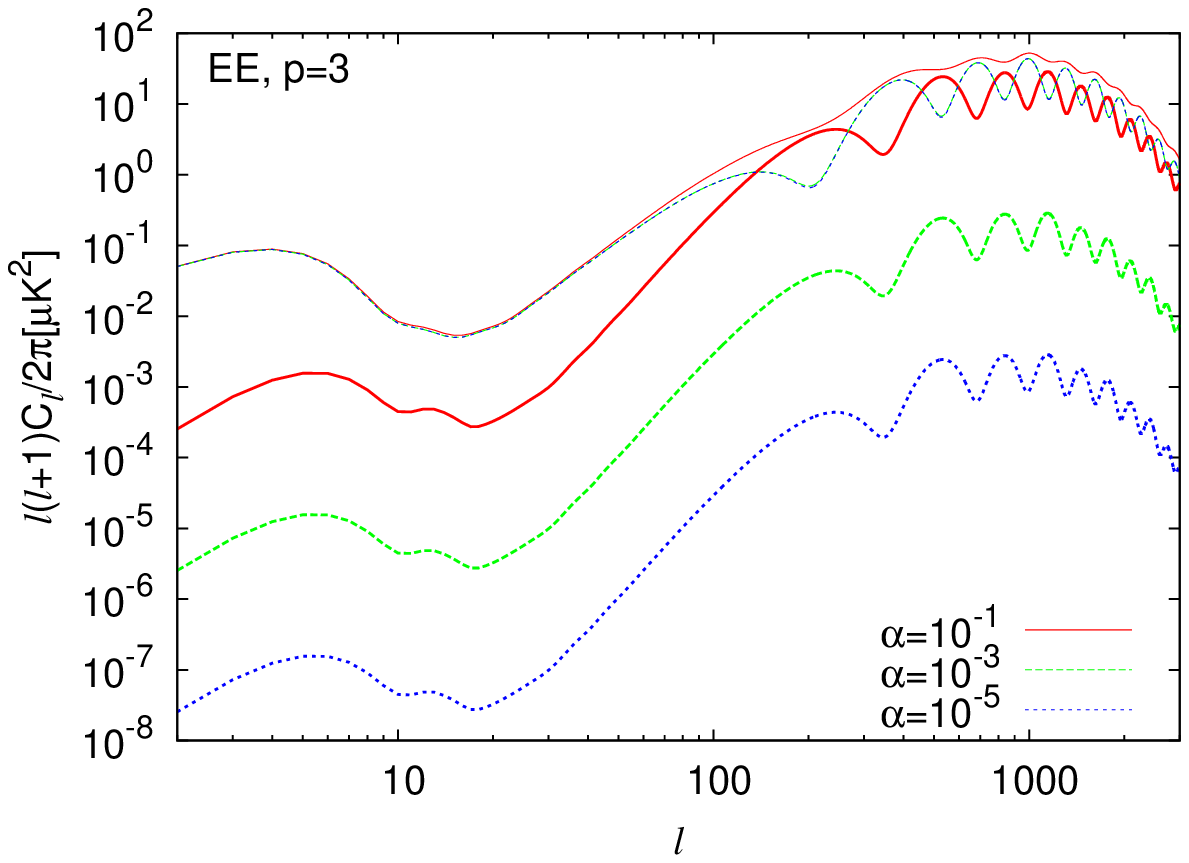}} &
    \hspace{-5mm}\scalebox{.4}{\includegraphics{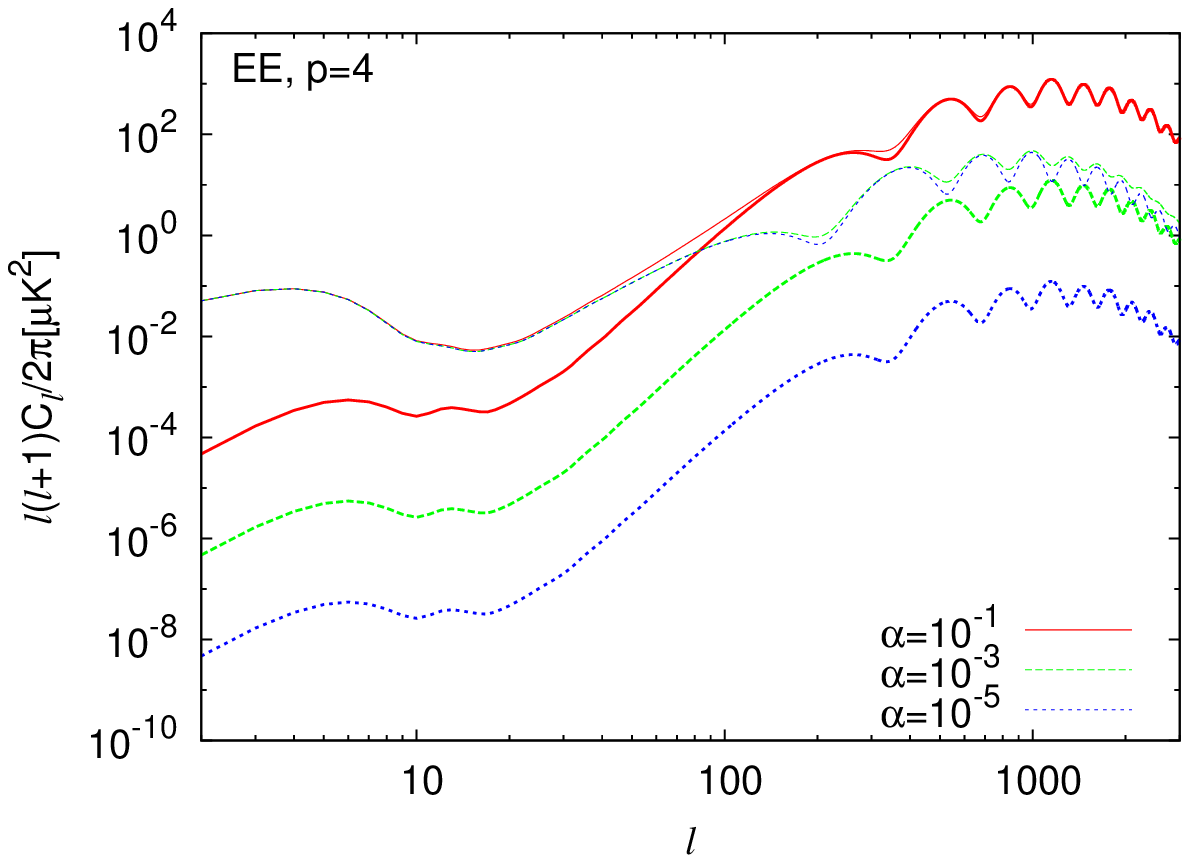}} \\
    \hspace{-5mm}\scalebox{.4}{\includegraphics{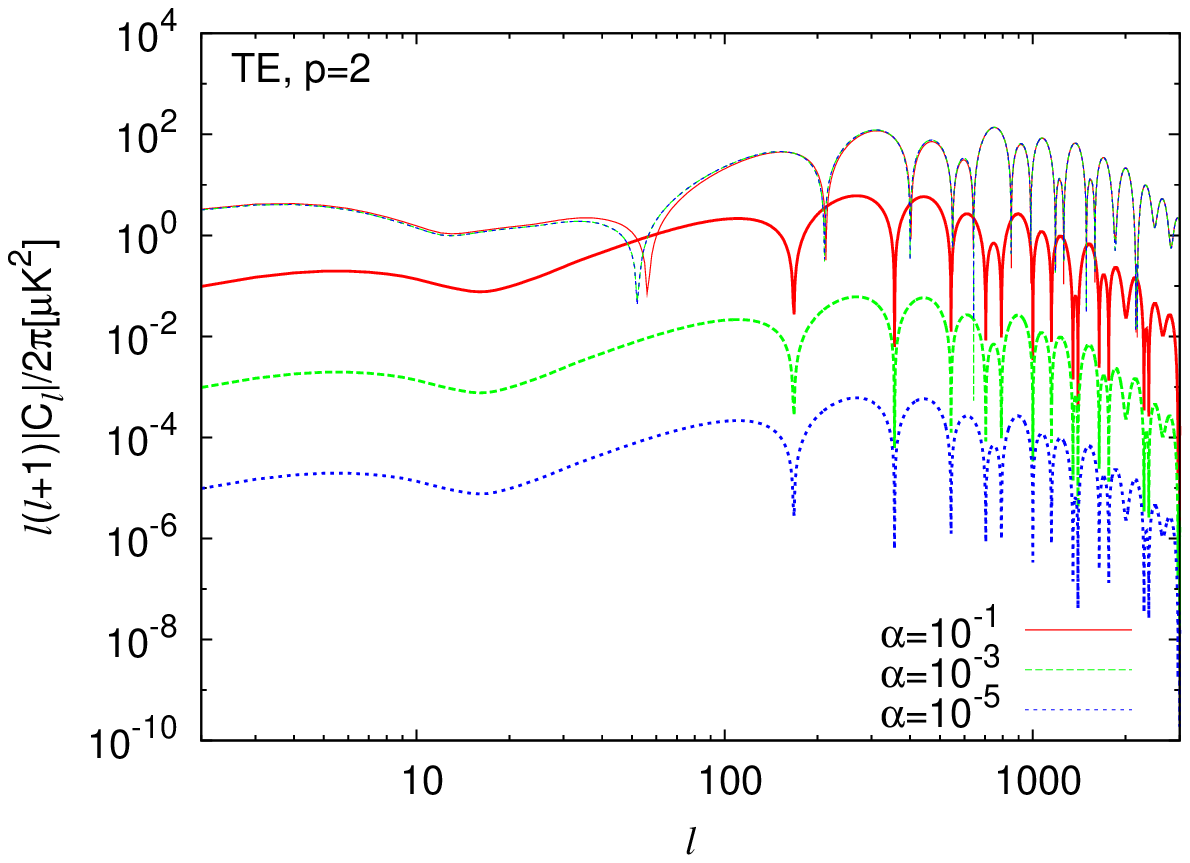}} &
    \hspace{-5mm}\scalebox{.4}{\includegraphics{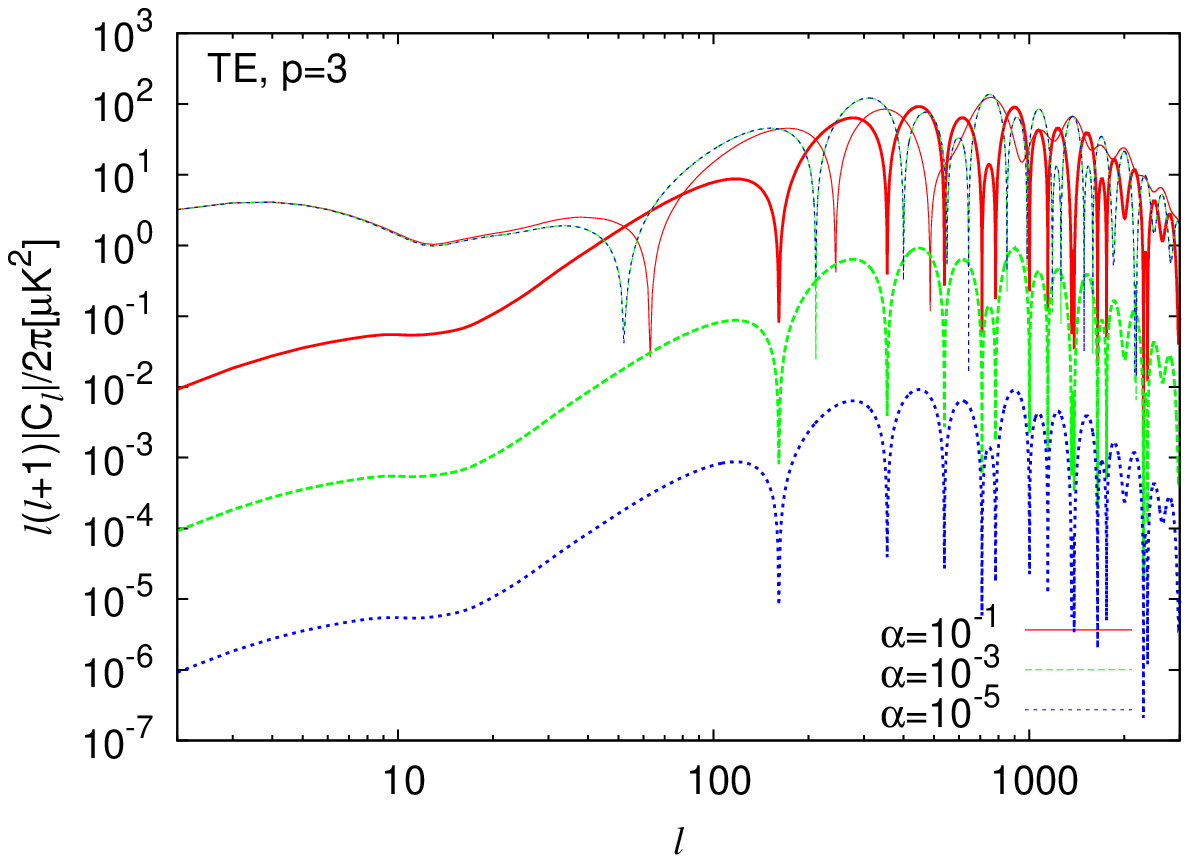}} &
    \hspace{-5mm}\scalebox{.4}{\includegraphics{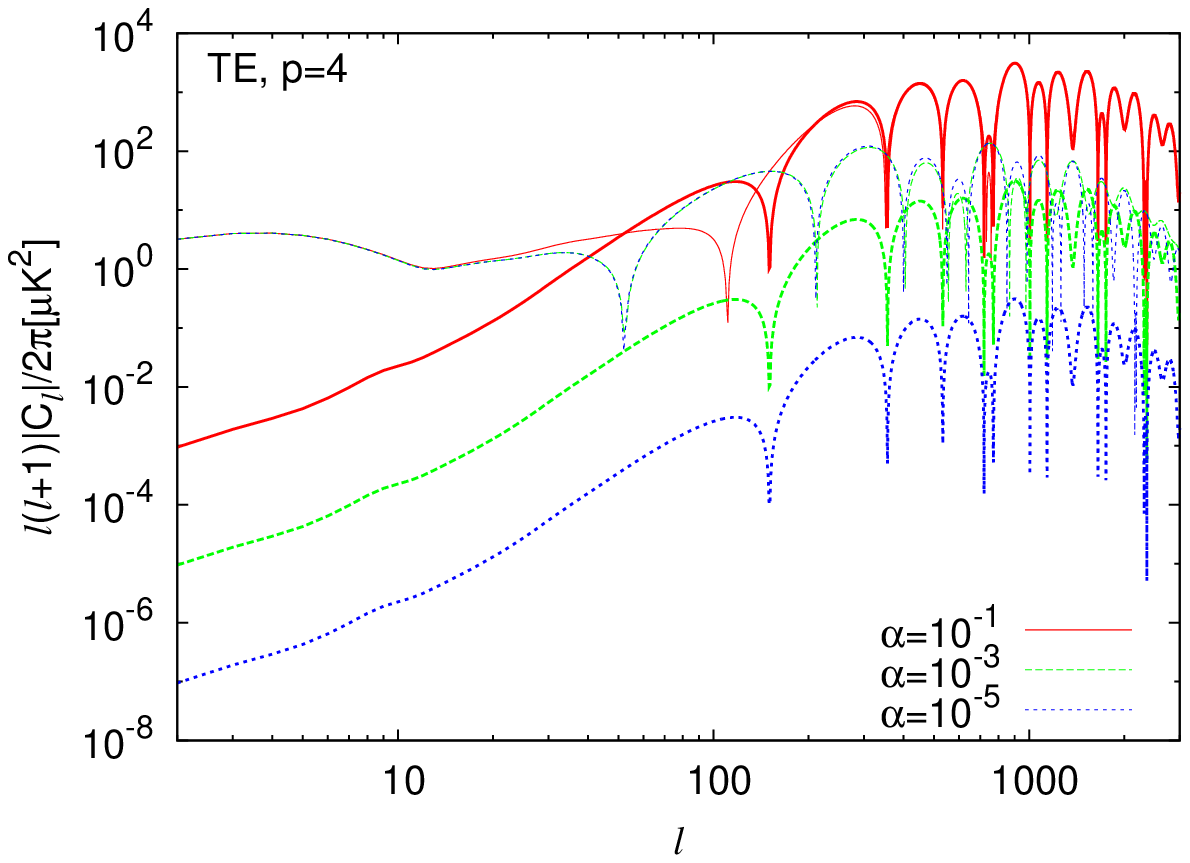}}     
  \end{tabular}
  \end{center}
  \caption{CMB power spectrum for different amplitudes and spectral indices of the isocurvature power spectrum. 
  In order from top to bottom, the temperature, E-mode polarization and their cross-correlation power spectra are plotted.
  Cases of $p=2$, 3, 4 are shown in order from left to right. In each panel, cases of $\alpha=10^{-1}$ (red solid), 
  $10^{-3}$ (green dashed), $10^{-5}$ (blue dotted) are plotted.
  Thick and thin lines show contributions from the isocurvature power spectrum alone and the sum of the adiabatic and isocurvature spectra, respectively.
  }
  \label{fig:cls}
\end{figure}

\begin{figure}
  \begin{center}
    \scalebox{1.}{\includegraphics{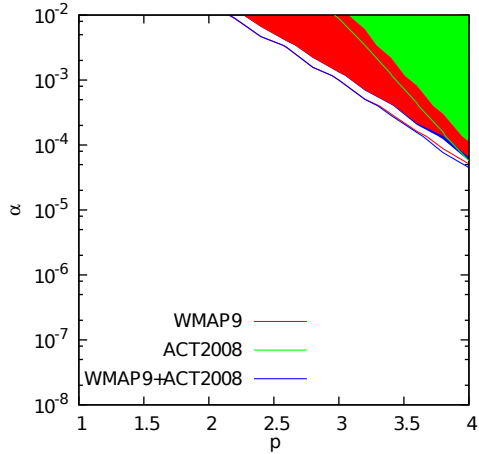}}
  \end{center}
  \caption{Constraints on the amplitude $\alpha$ and the spectral index $p$ of 
  the isocurvature power spectrum from CMB datasets.
  Results from WMAP9 (red) ACT2008 (green) and WMAP9+ACT2008 (blue) are shown 
  The contours and shaded regions respectively show parameter regions excluded at 1 and 2$\sigma$ levels.
  }
  \label{fig:loglike}
\end{figure}

First, we see to what extent an isocurvature power spectrum is constrained 
from measured shapes of the CMB power spectrum. 
In Fig.~\ref{fig:cls}, we plot the CMB power spectrum with several different values of the 
amplitude and spectral indices of the isocurvature power spectrum.
Fig.~\ref{fig:loglike} shows 1 and 2$\sigma$ constraints in the
$p$-$\alpha$ plane from the observed CMB power spectrum.
We adopt the likelihood functions from the WMAP 9-year results~\cite{Hinshaw:2012aka,Bennett:2012fp} 
and the ACT 2008 results~\cite{Das:2010ga,Dunkley:2010ge}.
From the figure, we can see that for $p\lesssim 3$, WMAP is more sensitive than ACT.
On the other hand, as the spectral index becomes larger $p\gtrsim4$, 
the CMB spectrum at higher multipoles $\ell\gtrsim1000$, which
can be measured by ACT, becomes more sensitive. 

We however should note that here we have fixed other cosmological parameters and in addition 
contributions of foregrounds components that include the thermal Sunyaev-Zel'dovich effect 
and point-sources.  Varying these  may non-negligibly loosen the constraints, 
in particular for ACT, which observes small angular scales
where these foregrounds dominate primary anisotropies.
In effect, our constraints are at least an order of magnitude stronger than 
the constraints presented by Planck~\cite{Ade:2013uln}, although it is difficult to compare
these with our constraints in a straightforward way, 
as the parameterizations for the isocurvature power spectrum are different\footnote{
In Ref.~\cite{Ade:2013uln}, constraints on the isocurvature power spectrum with variable spectral index are not given 
for uncorrelated cases, but only for generally correlated cases. This could be another reason for not comparing the constraints
with ours in a straightforward way. However, since upper bounds on the amplitude $\alpha$ tends to be least stringent
for uncorrelated cases (explicitly shown in  Ref.~\cite{Ade:2013uln} in cases of a nearly scale-invariant isocurvature power spectrum),
we expect that constraints on $\alpha$ from the Planck result for uncorrelated cases would not differ significantly from generally correlated cases.}.
Our constraints here should be regarded as very optimistic.

\section{Mass function}
\label{sec:massf}

Since the CMB power spectrum can measure primordial perturbations
only on large scales, isocurvature perturbations are allowed
to have large amplitudes on small scales if the power spectrum is blue-tilted.
Such isocurvature perturbations with a blue-tilted spectrum can significantly
affect structure formation on small scales and can increase the number of 
dark matter halos in the high-redshift universe.
To investigate this effect, we evaluate the mass function of dark matter halos 
for different isocurvature parameters in this section.

According to the Press-Schechter formalism~\cite{Press:1973iz},
the mass function of dark matter halos is given by
\begin{equation}
\frac{dn}{d\ln M}=\bar\rho_m f(\nu)\frac{d\nu}{dM}, 
\end{equation}
where $\bar \rho_m$ is the mean energy density of matter,
$f(\nu)$ is the occupation function and 
we define $\nu\equiv (\delta_{\rm cr}/\sigma(M))^2$ with
$\delta_{\rm cr}$ and $\sigma(M)$ respectively being the critical overdensity and 
the variance of matter fluctuations in a sphere of radius $R$ satisfying 
$M=\frac{4\pi}3\bar\rho_m R^3$. 

Given a matter power spectrum $P(k)$, 
$\sigma(M)$ is obtained by 
\begin{equation} 
\sigma(M)^{2}= \int\frac{k^{2}dk}{2\pi^{2}}
W^{2}(kR)P(k),
\label{sigma8}
\end{equation}
where $W(kR)$ is a top-hat window function with radius $R$, 
\begin{equation} 
W(kR)\equiv \dfrac{3}{\left( kR \right)^{3}}
\left(\sin(kR)-kR\cos (kR) \right).
\end{equation}
Regarding the occupation function $f(\nu)$, we adopt the one proposed by Sheth and Tormen~\cite{Sheth:1999mn},
\begin{equation}
\nu f(\nu)=A(a)\left(1+(q\nu)^{-a}\right)\left(\frac{q\nu}{2\pi}\right)^{1/2}\exp\left[-\frac{q\nu}2\right], 
\label{eq:occ}
\end{equation}
where $a=0.3$, $q=0.75$ and $A(a)$ is the normalization so that occupation function satisfies $\int d\nu f(\nu)=1$
or, equivalently, $\int dM \frac{dn}{d\ln M}=\bar \rho_m$.

\begin{figure}
  \begin{center}
  \begin{tabular}{cc}
    \hspace{-5mm}\scalebox{.6}{\includegraphics{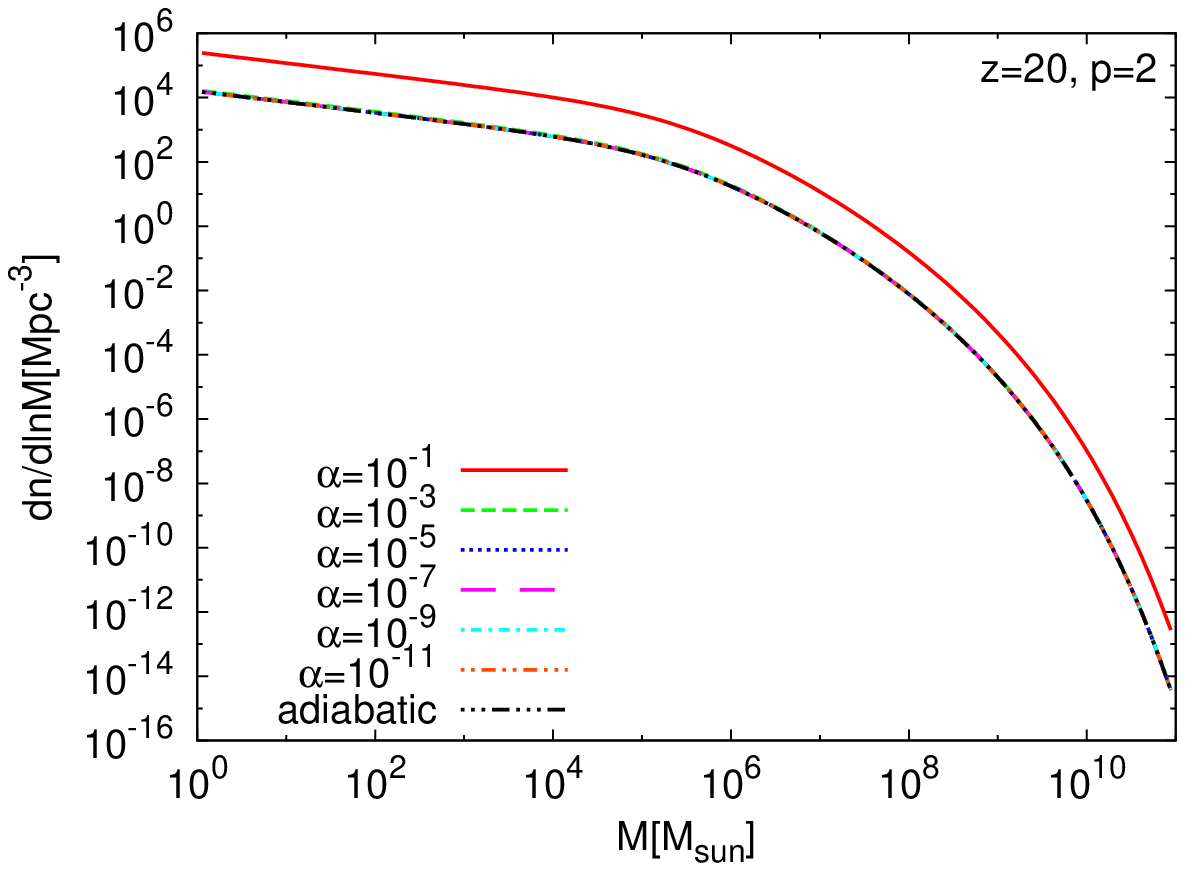}} &
    \hspace{-5mm}\scalebox{.6}{\includegraphics{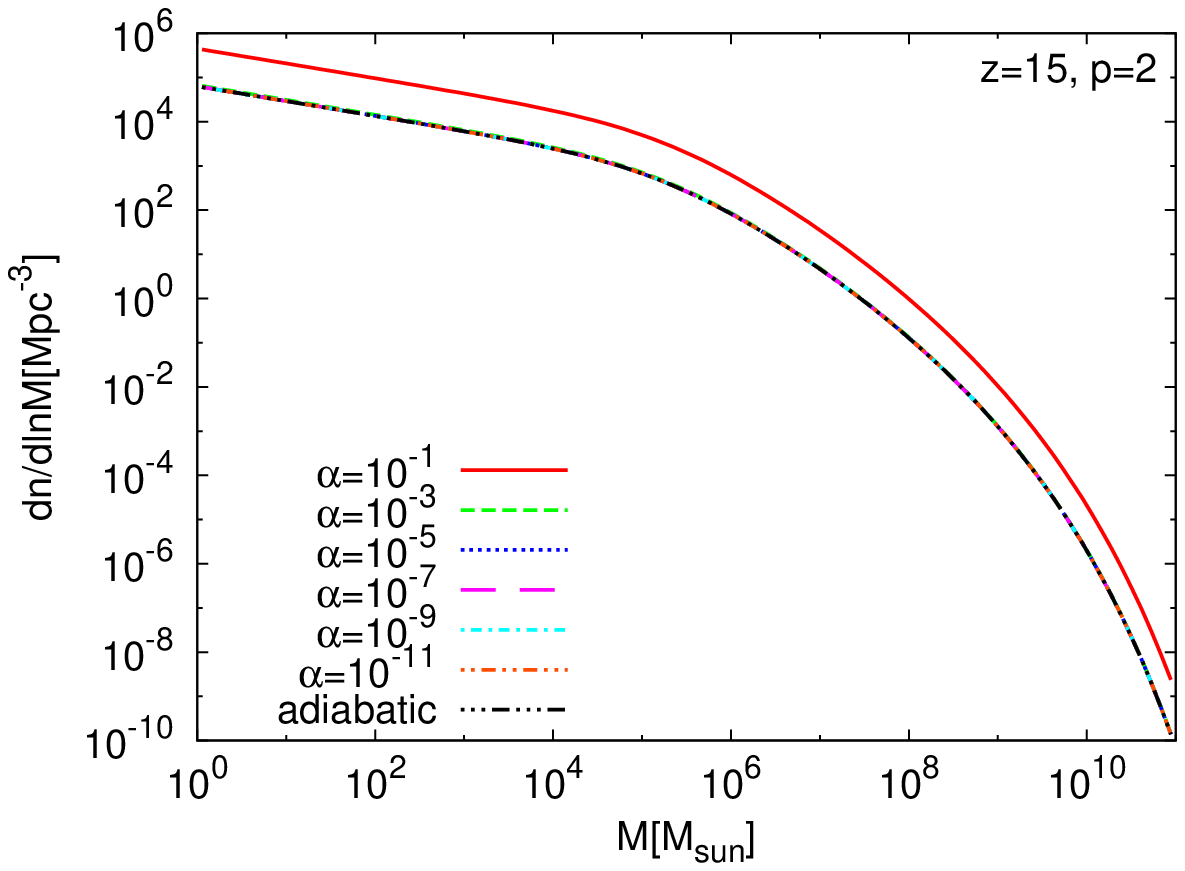}} \\
    \hspace{-5mm}\scalebox{.6}{\includegraphics{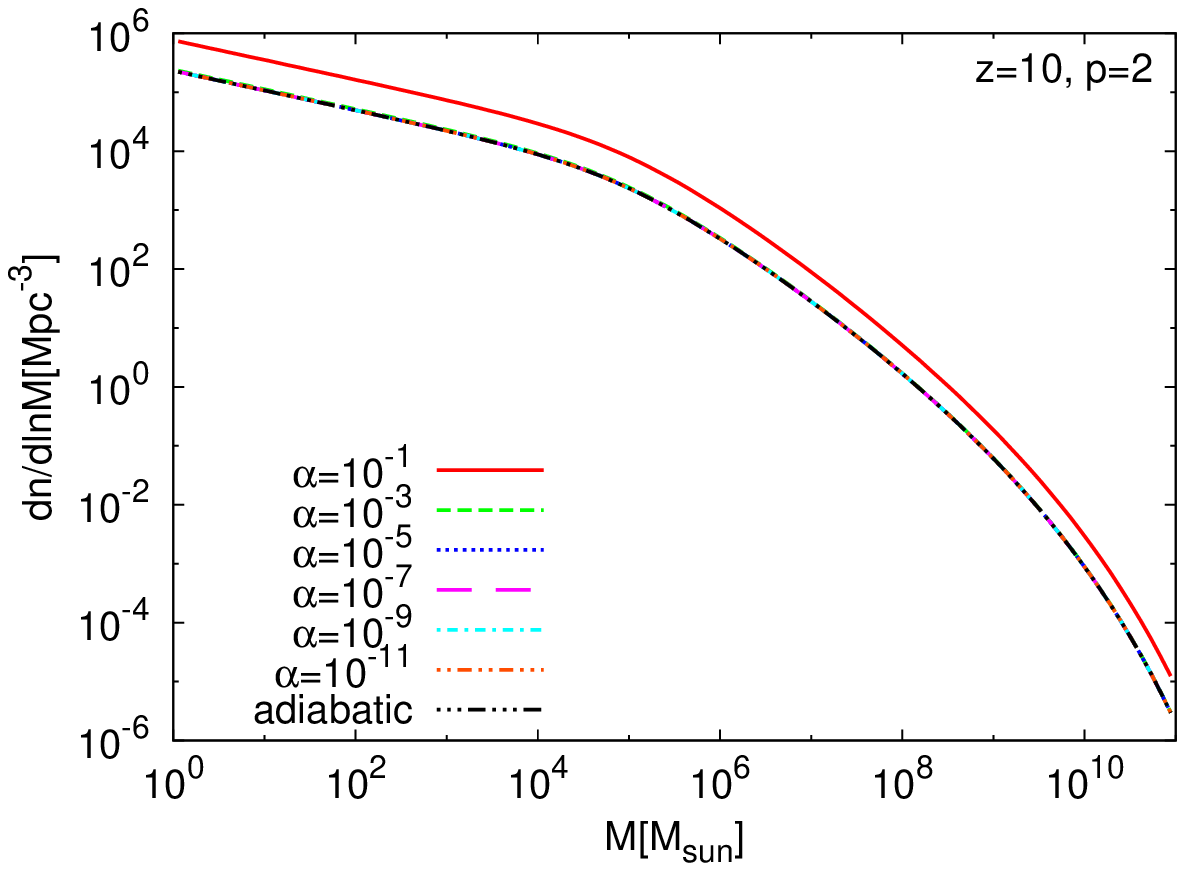}} &
    \hspace{-5mm}\scalebox{.6}{\includegraphics{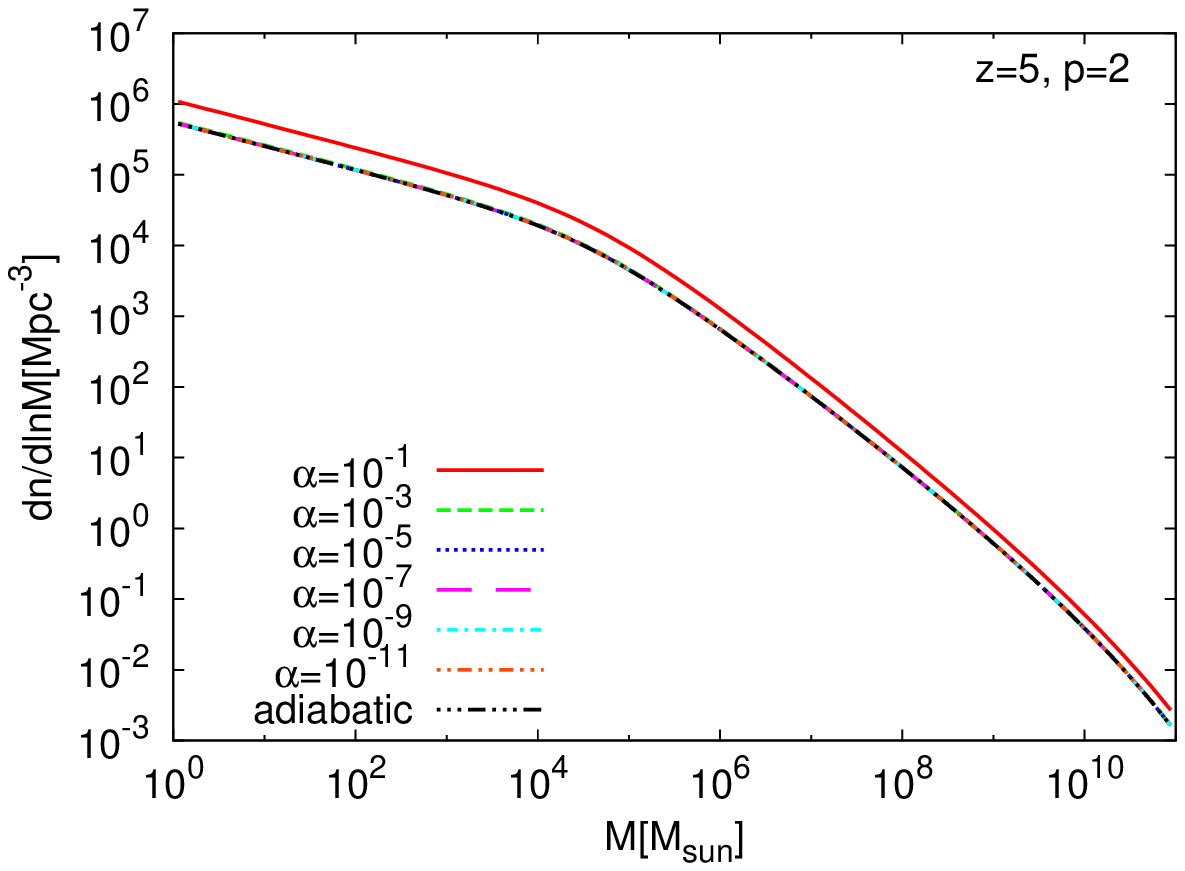}} 
  \end{tabular}
  \end{center}
  \caption{Mass function $dn/d\ln M$ for the case of isocurvature power spectrum with $p=2$. In order from top-left to bottom-right, 
  the panels show mass functions at redshifts $z$=20, 15, 10, 5.  Each line corresponds to a different value of the 
  amplitude of isocurvature power spectrum $\alpha=0.1$ (red solid), $1\times10^{-3}$ (green short-dashed), $1\times10^{-5}$ (blue dotted), 
  $1\times10^{-7}$ (magenta long-dashed), $1\times10^{-9}$ (light blue dot-dashed), $1\times10^{-11}$ (orange two-dot-chain).
  As a reference, the pure adiabatic case is also plotted (black three-dot-chain).
  }
  \label{fig:p2}
\end{figure}

\begin{figure}
  \begin{center}
  \begin{tabular}{cc}
    \hspace{-5mm}\scalebox{.6}{\includegraphics{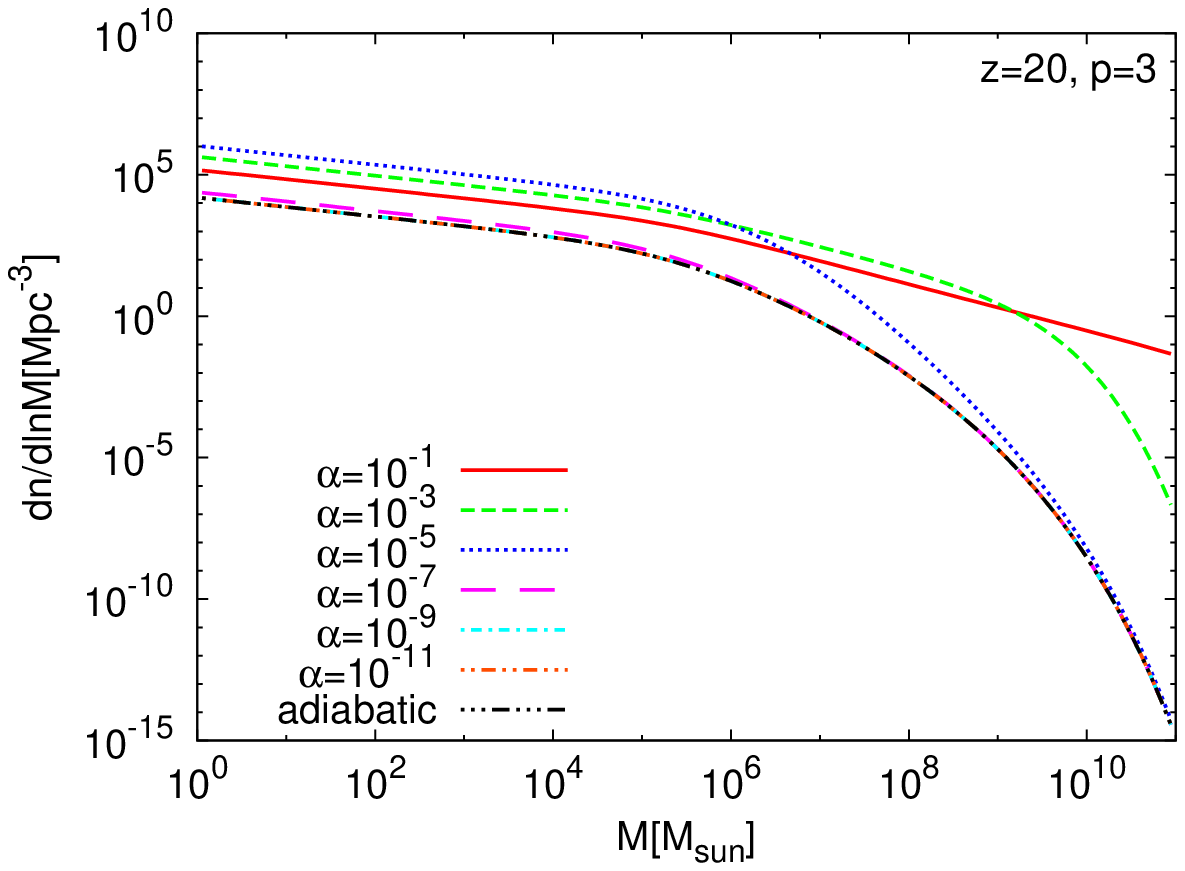}} &
    \hspace{-5mm}\scalebox{.6}{\includegraphics{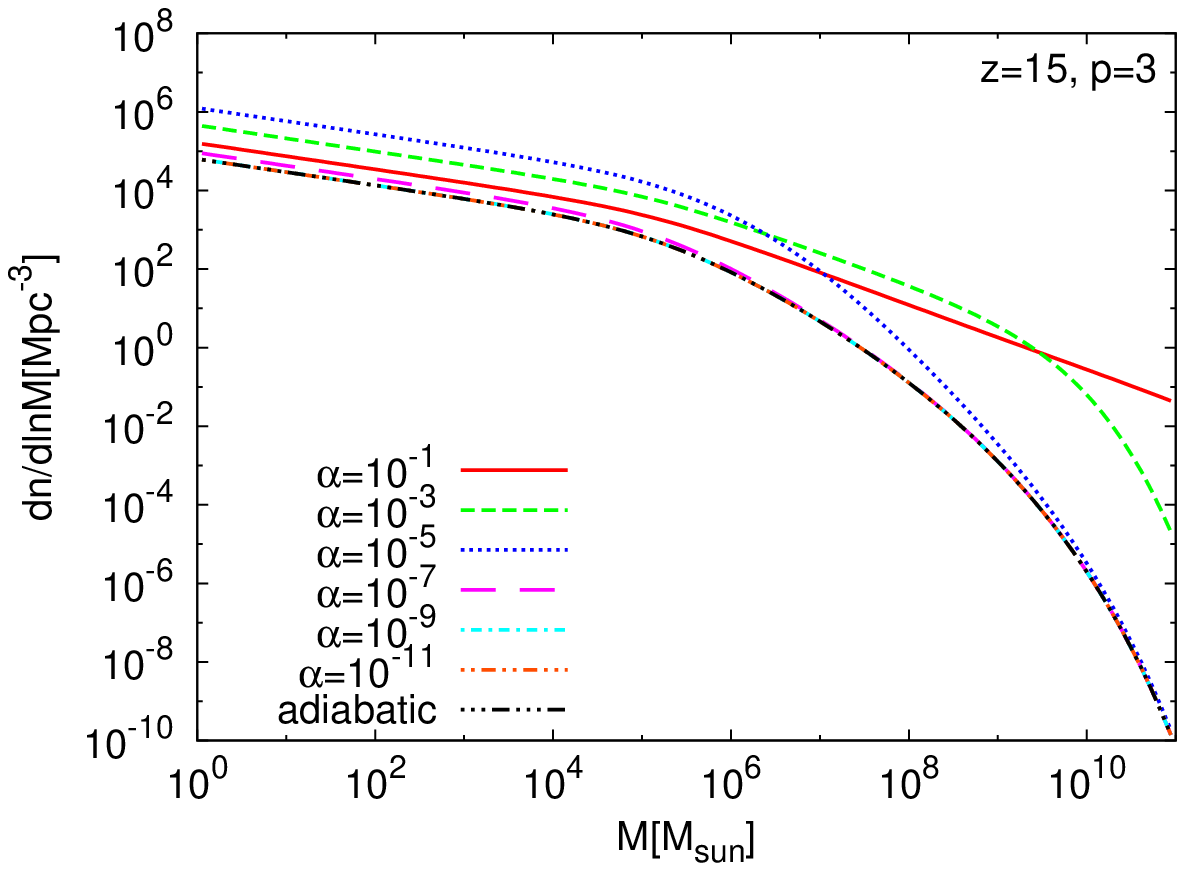}} \\
    \hspace{-5mm}\scalebox{.6}{\includegraphics{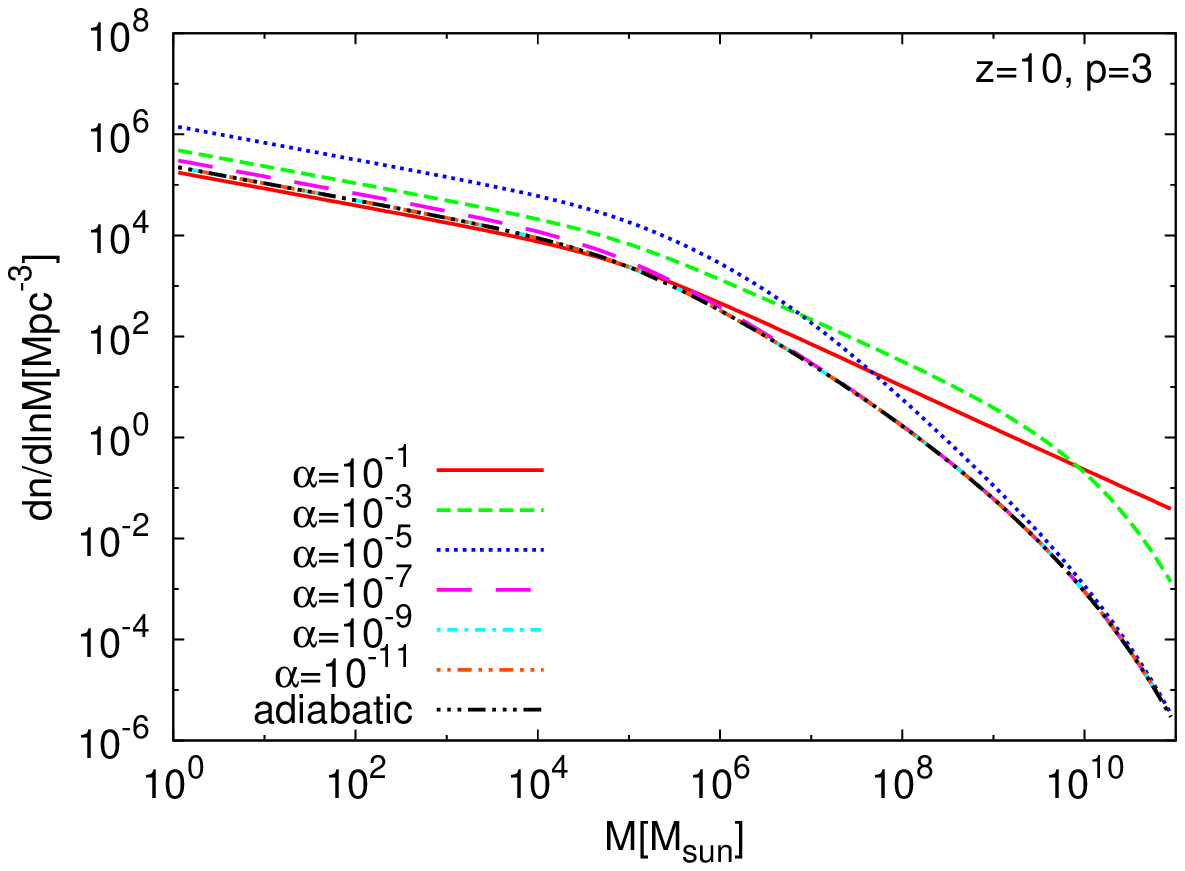}} &
    \hspace{-5mm}\scalebox{.6}{\includegraphics{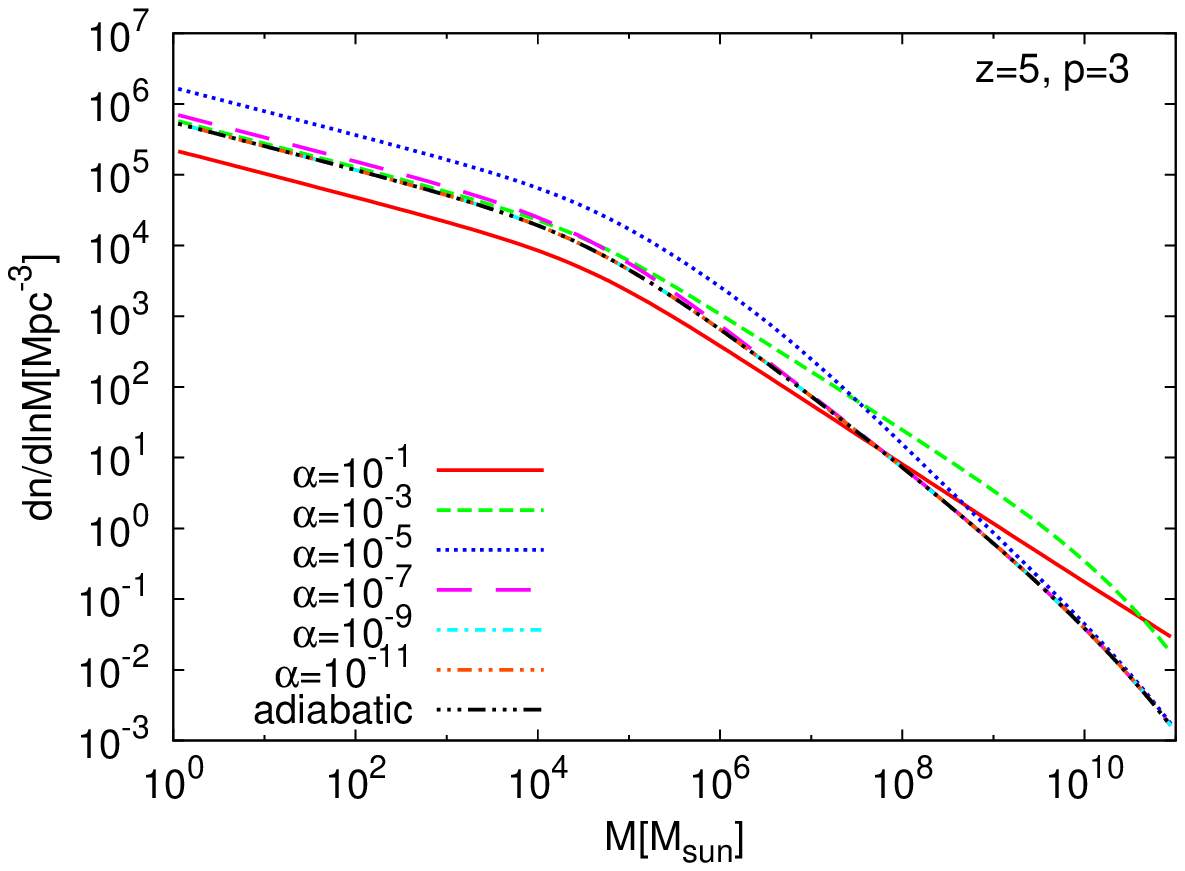}} 
  \end{tabular}
  \end{center}
  \caption{Same as in Fig.~\ref{fig:p2} but for $p=3$.
  }
  \label{fig:p3}
\end{figure}

\begin{figure}
  \begin{center}
  \begin{tabular}{cc}
    \hspace{-5mm}\scalebox{.6}{\includegraphics{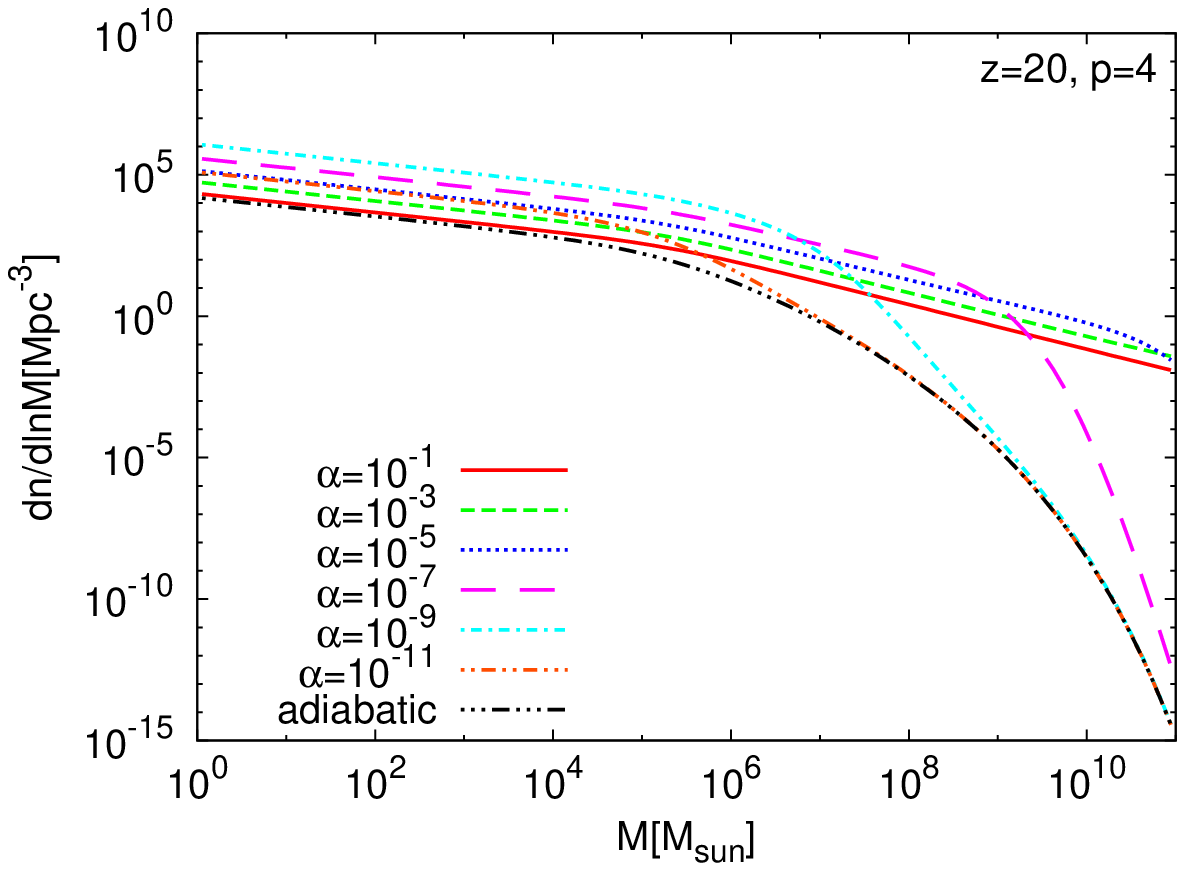}} &
    \hspace{-5mm}\scalebox{.6}{\includegraphics{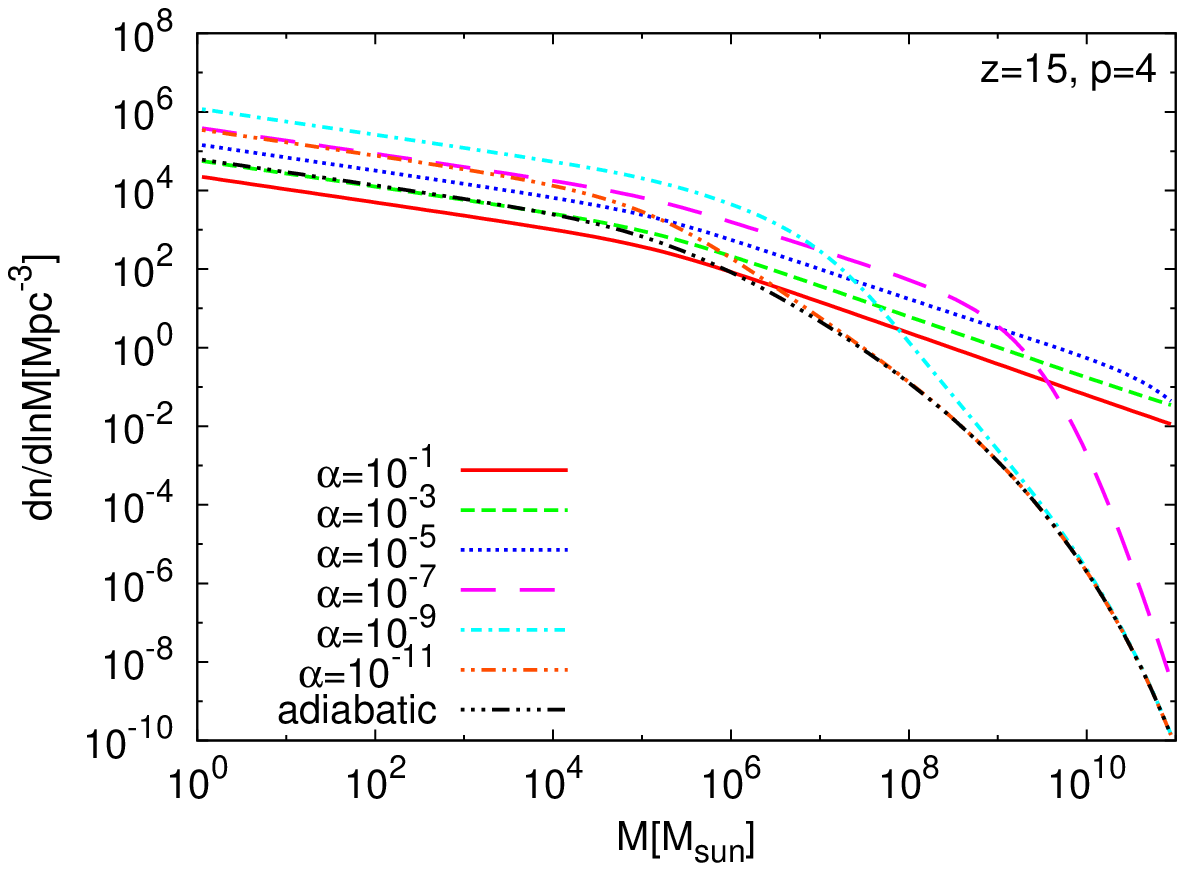}} \\
    \hspace{-5mm}\scalebox{.6}{\includegraphics{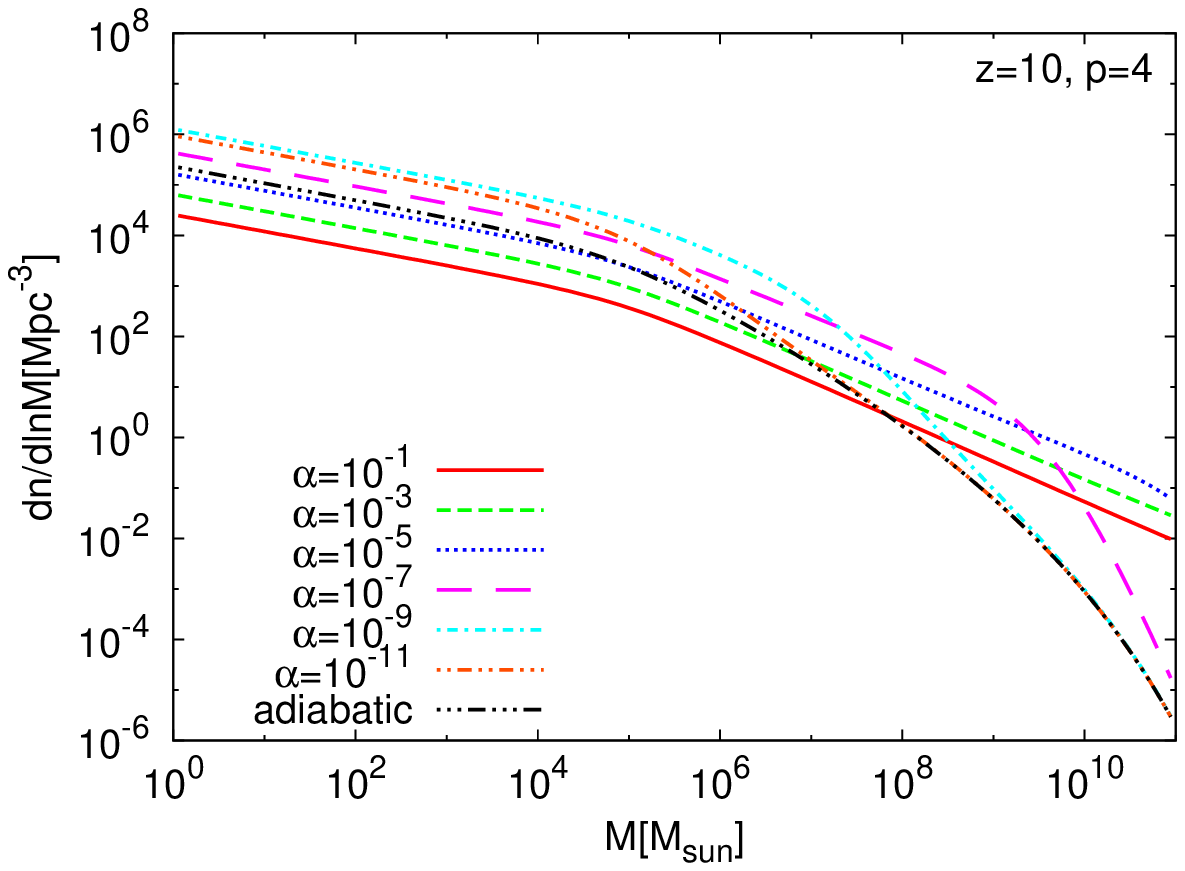}} &
    \hspace{-5mm}\scalebox{.6}{\includegraphics{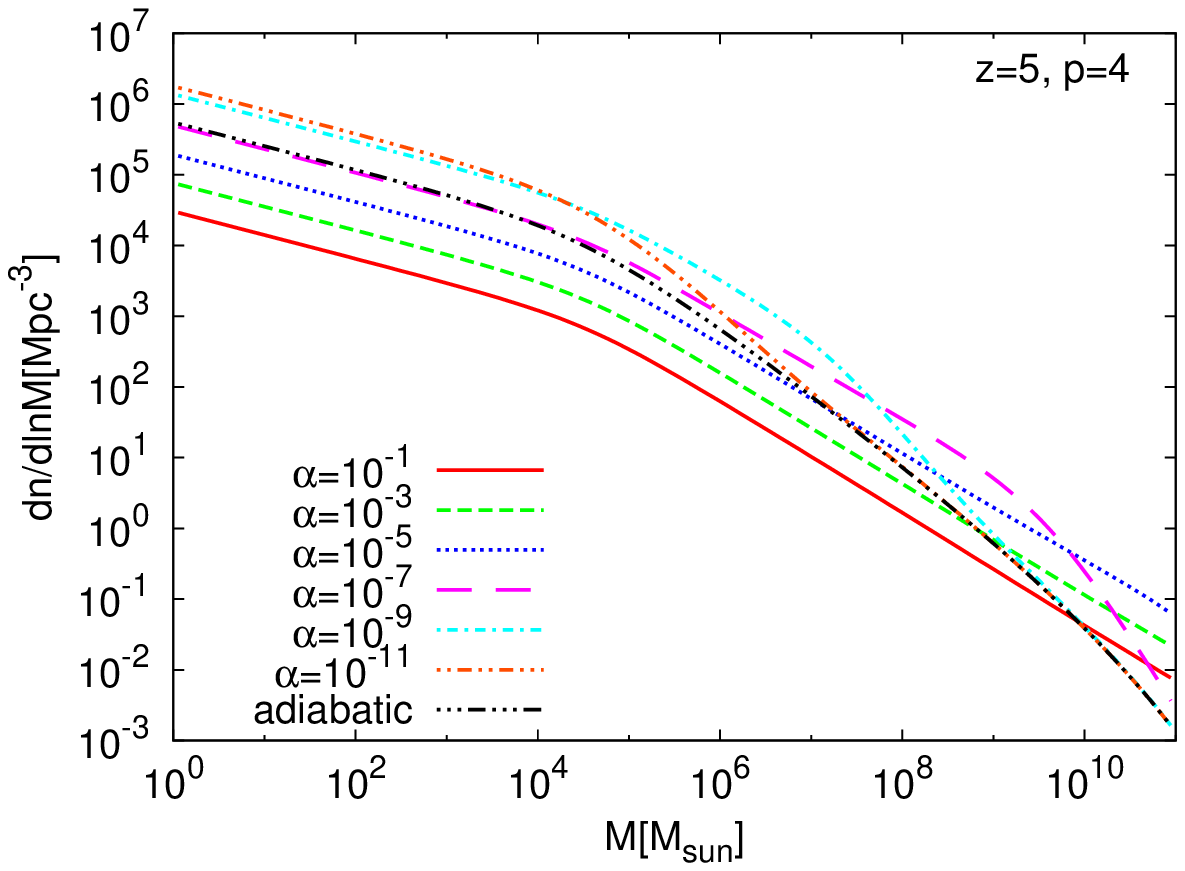}} 
  \end{tabular}
  \end{center}
  \caption{Same as in Fig.~\ref{fig:p2} but for $p=4$.
  }
  \label{fig:p4}
\end{figure}

In Figs.~\ref{fig:p2}-\ref{fig:p4}, we plot the mass function for various isocurvature power spectra
specified by parameter sets $(\alpha, p)$. In these figures, 
$p$ is varied from $2$ to $4$. In each figure, 
mass functions at different redshifts $z=20, 16, 12, 8, 4$ are plotted from the top left panel
to the bottom right panel. In each panel, 
different lines correspond to different amplitudes of the isocurvature power spectrum $\alpha$.

From Fig.~\ref{fig:p2}, 
we can see that, for moderately blue-tilted isocurvature power spectra with $p=2$,
the effects of isocurvature perturbations
can be seen only for large amplitude $\alpha\gtrsim10^{-1}$;
for smaller $\alpha$ the mass function can hardly be distinguished from that for 
the vanilla model (the purely adiabatic case).
For $\alpha\gtrsim10^{-1}$, 
the presence of the isocurvature power spectrum amplifies the mass
function independently of mass in the range $[1M_\odot,10^{11}M_\odot]$. 
On the other hand, from Figs.~\ref{fig:p3} and \ref{fig:p4}, which show the effects of more blue-tilted 
isocurvature spectra with $p=3$ and $p=4$, we can see qualitative differences from those for $p=2$.
At larger masses $M\to 10^{11}M_\odot$, we see that the mass function is enhanced 
monotonically as  $\alpha$ increases, which is also true for $p=2$ as already seen above. 
However, at the lower mass end $M\to 1M_\odot$, the effects on the mass function become more 
complicated: as $\alpha$ increases from zero, the mass function first
increases and, at some point in $alpha$,  
it tends  to decrease. At smaller redshift and/or with larger $p$, the mass function at $M=1M_\odot$ 
can become even smaller than that of the vanilla model with $\alpha=0$. 

This somewhat counterintuitive behavior of the mass function comes from the normalization 
$\int dM \frac{dn}{d\ln M}=\bar \rho_m$.
To see this, it is convenient to define a critical mass $M_{\rm cr}$ as 
\begin{equation}
\sigma(M_{\rm cr})=\delta_{\rm cr}.
\end{equation}
From Eq.~\eqref{eq:occ} as well as Figs.~\ref{fig:p2}-\ref{fig:p4}, one can see that halos with masses up 
to $M_{\rm cr}$ or equivalently $\nu=1$ can be abundantly produced while the number of halos with 
larger masses are suppressed exponentially. In particular, Figs.~\ref{fig:p3} and \ref{fig:p4} show that the
presence of large isocurvature perturbations with  blue spectra makes $M_{\rm cr}$ much larger than  in the 
pure adiabatic vanilla model. Given fixed total mass in the universe $\bar \rho_m=\int dM \frac{dn}{d\ln M}$, any
increase in the number of halos with large masses should be compensated with a decrease in those with small masses.

However, the above explanation may not be satisfactory enough
as it lacks a physical pictures, which we try to describe below.
First, we divide space into cells with comoving volume $V_1$.
We also consider the division of space into larger cells with $V_2$~($>V_1$). 
The variance of matter fluctuations for $V_1$ and $V_2$ is given
by $\sigma(M_1)$ and $\sigma(M_2)$ with the averaged mass $M_1 = \bar \rho_m V_1$ and
$M_2 = \bar \rho_m V_2$ for $V_1$ and $V_2$, respectively.

Let us consider the case for which, while $\sigma (M_1) > \delta_{\rm cr}$,
 $\sigma (M_2) < \delta _{\rm cr}$~(low density fluctuation case).
When we divide space into cells with $V_1$,
the actual density fluctuations in a cell are larger than
$\delta_{\rm cr}$ in some cells, and dark halos are formed in such cells.
For the division into larger cells, the actual density fluctuations in a
large cell $V_2$ cannot exceed $\delta_c$. Therefore, large cells cannot
collapse to dark halos~(as discussed above, the number destiny of halos for $M_2$ are
suppressed exponentially).

Suppose that both $\sigma (M_1)$ and $\sigma (M_2)$ are larger than
$\delta_{\rm cr}$~(large density fluctuation case).
Dividing space into small cells with $V_1$,
we still find that actual density fluctuations in a cell are larger than
$\delta_{\rm cr}$ in some cells, and dark matter halos are formed in such cells.
We consider a large cell with $V_2$ which includes such small cells.
Since $\sigma (M_2) > \delta_c$, this large cell can have density fluctuations larger than $\delta_{\rm cr}$ and
a large halo is formed in this cell. In the formation of a large halo,
small halos in the cell are merged into a large halo.

Therefore, as the density fluctuations are amplified, the number density
of large halos increases.
On the other hand, the number density of small halos decreases because
small halos merge to form a large halo (or due to the normalization as
discussed above).
As we will see in Section~\ref{sec:21cm}, this specific response of the mass function to $\alpha$
at relatively small masses $M\lesssim10^7M_\odot$ leads to nontrivial behavior in 21~cm fluctuations.


\section{Optical depth}
\label{sec:opt}

As we have seen in the previous section, the presence of a blue-tilted isocurvature power spectrum
significantly alters the number density of halos at high redshifts. 
In particular, the formation of massive halos is enhanced, which results in an
increase in the number density of ionizing sources, or galaxies, at this epoch.
This affects the reionization history of the universe,  to be discussed in this section.

To see how the isocurvature power spectrum affects the reionization history, we here follow the analysis
of Ref. ~\cite{Tegmark:1993dg}. The ionized fraction of IGM, $\chi$, should be given by
\begin{equation}
\chi=\min[f_{\rm s}f_{\rm uvpp}f_{\rm ion},~1], 
\label{eq:chi}
\end{equation}
where $f_{\rm s}$ is the fraction of matter in the collapsed objects which can host galaxies emitting ionizing UV photons,
$f_{\rm uvpp}$ is the number of UV photons emitted into the IGM per
proton in the collapsed objects, and $f_{\rm ion}$ is the
net efficiency of ionization by a single UV photon.
Noting that a fraction 0.0073 of the rest mass is released in stellar 
burning of hydrogen to helium, 
$f_{\rm uvpp}$ can be further decomposed into 
\begin{equation}
f_{\rm uvpp}=0.0073\left(\frac{m_pc^2}{13.6 {\rm eV}}\right)(1-Y_p)f_{\rm burn}f_{\rm uv}f_{\rm esc},
\end{equation}
where $m_p$ is the proton mass,
$1-Y_p\simeq0.75$ is the mass fraction of hydrogen in IGM, $f_{\rm burn}$ is the mass fraction of hydrogen burned, 
$f_{\rm uv}$ is the fraction of energy released as UV photons, and $f_{\rm esc}$ is the fraction of UV photons that escape from galaxies.
To compile parameters whose values are not precisely known, we define a parameter 
\begin{equation}
f_{\rm net}=f_{\rm burn}f_{\rm uv}f_{\rm esc}f_{\rm ion}, 
\end{equation}
which has a fairly large uncertainty. 
Then Eq.~\eqref{eq:chi} can be rewritten as 
\begin{equation}
\chi=\min[3.8\times 10^5 f_{\rm net} f_{\rm s},~1].
\end{equation}

The fraction of matter in collapsed objects $f_{\rm s}$ should be given by
\begin{equation}
f_s(z)\equiv\frac1{\bar\rho_m(z)}\int^\infty_{M_*(z)} \frac{dn}{d\ln M}(M,z) dM, 
\end{equation}
where $M_*(z)$ corresponds to the minimum mass of collapsed objects whose virial temperature should be
larger than $10{^4}$K, above which  atomic cooling becomes effective. 
According to Ref.~\cite{Iliev:2002gj}, $M_*(z)$ can be approximately given as
\begin{equation}
M_*(z)=3.95\times 10^7\left(\frac{\Omega_mh^2}{0.15}\right)^{-1}
\left(\frac{1+z}{10}\right)^{-3/2}
M_\odot.
\label{eq:m*}
\end{equation}

We note that similar calculation of reionization history is done in Ref. \cite{Sugiyama:2003tc}
for a blue-tilted isocurvature power spectrum. In particular for $f_{\rm net}=10^{-4}$, 
the reionization model we adopted here is almost the same as the model for POP II star in the reference.
In this case, we indeed confirm that evolution of the reionization fraction in our calculation show a good agreement with 
Ref. \cite{Sugiyama:2003tc}.

The optical depth of reionization is given by
\begin{equation}
\tau_{\rm reion}=\int dt ~\chi(t) \bar n_p(t)\sigma_{\rm T}, 
\end{equation}
where $\bar n_p(t)$ and $\sigma_{\rm T}$ are the mean number density of protons and the  cross-section for Thomson scattering.

\begin{figure}
  \begin{center}
    \scalebox{1.}{\includegraphics{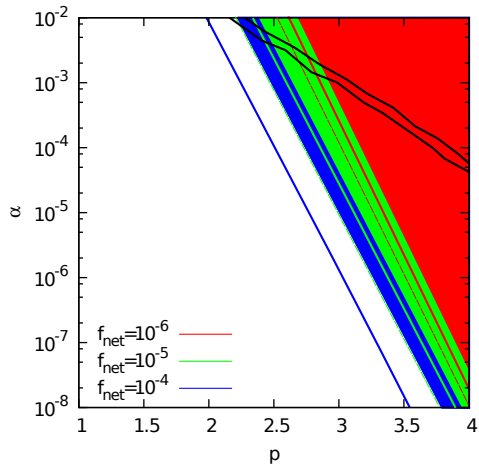}}
  \end{center}
  \caption{Constraints on the isocurvature power spectrum from considerations on reionization.
  Here we take the efficiency parameter for reionization $f_{\rm net}$ to $10^{-6}$ (red), $10^{-5}$ (green) and $10^{-4}$ (blue), 
  and show contours of the reionization optical depth $\tau_{\rm reion}$ for each $f_{\rm net}$.
  For each value of $f_{\rm net}$, shaded region shows parameter regions which give $\tau_{\rm reion}\ge0.13$, while 
  thick solid and thin dashed lines respectively show parameter regions which give $\tau_{\rm reion}=0.09$ and $0.05$.
  Note that for $f_{\rm net}=10^{-4}$, $\tau_{\rm reion}$ exceeds 0.05 even if the isocurvature power spectrum vanishes (i.e. $\alpha=0$), 
  so that the blue thin dashed line does not appear in the plot.
  As reference, 1 and 2$\sigma$ constraints from the CMB dataset (WMAP9+ACT2008) shown in Fig.~\ref{fig:loglike} are also plotted (black solid line).
  }
  \label{fig:opt}
\end{figure}

Fig.~\ref{fig:opt} shows the contour plot of the optical depth $\tau_{\rm reion}$ in the $p$-$\alpha$ plain and constraints on these parameters
from the WMAP9 estimate of the optical depth $\tau_{\rm reion}=0.09\pm 0.04$ (95\% C.L.)~\cite{Hinshaw:2012aka}\footnote{
Although this constraint on $\tau_{\rm reion}$  is obtained in the absence of isocurvature perturbations, it is still valid in cases with 
non-vanishing isocurvature perturbations. This is because $\tau_{\rm reion}$ is constrained by the ``reionization bump"
in the CMB polarization spectrum at low multipoles $\ell \sim 10$ which a blue-tilted isocurvature power spectrum would 
not affect significantly as shown in Fig.~\ref{fig:cls}.
}. The lowest value of the reionization efficiency $f_{\rm net}=10^{-6}$ we take here should be regarded as a case for very 
inefficient reionization following the discussion of Ref.~\cite{Tegmark:1993dg}.

From the figure, we can see that optical depth of reionization is sensitive to highly blue-tilted isocurvature spectrum with $p\gtrsim3$.
Even if we assume a very low value of $f_{\rm net}=10^{-6}$, $\alpha$ should not be large for $p\gtrsim3$.
We found that WMAP constraints on $\tau_{\rm reion}<0.13$ (95\% C.L.) leads to a constraint
\begin{equation}
\log_{10} \alpha<-4.2p-1.1\log_{10}f_{\rm net}-3.0, 
\end{equation}
which is valid for $10^{-6}\le f_{\rm net}\le10^{-4}$.
Compared with the constraints directly obtained from the CMB power
spectrum from WMAP9+ACT2008, Fig.~\ref{fig:opt}
shows that considerations of reionization history can give strong 
constraints on blue-tilted isocurvature 
power spectrum for large spectral indices $p\gtrsim3$ despite the large uncertainties in reionization efficiency $f_{\rm net}$.

\section{21cm fluctuations}
\label{sec:21cm}

Here we discuss signatures of a blue-tilted isocurvature power spectrum on redshifted 21cm line fluctuations.
As we discussed in the previous section, a blue-tilted isocurvature
power spectrum induces small-mass halos.
Among them, we mainly focus on ``minihalos''  whose mass is too small
to host a galaxy.
Although minihalos are not considered to be luminous objects,
a minihalo can make observable 21~cm line signals~\cite{Iliev:2002gj,Furlanetto:2002ng}.
Our analysis is basically the same as those in Refs.~\cite{Iliev:2002gj,
Chongchitnan:2012we}. As assumed in these references, we also 
assume that minihalos are modeled as a ``truncated isothermal sphere" 
with physical radius $r_t(M)$, gas temperature $T_K(l,M)$, dark matter velocity dispersion 
$\sigma_{\rm V}(l,M)$ and density profile $\rho(l,M)$, where $l$ is the distance from the center of a minihalo.
We denote the frequency of 21cm line emission in the rest frame as $\nu_0=1.4$ GHz.
We also note that in this section, any parameter dependences of quantities are significantly abbreviated.

Let us consider the brightness temperature of 
photons with frequency $\nu'$ at redshift $z$ 
which penetrate a minihalo of mass $M$ with an impact parameter $r$.
The brightness temperature is given by
\begin{equation}
T_b(\nu',z,r,M)=T_{\rm CMB}(z)e^{-\tau_{\rm 21cm}}
+\int dR~T_{\rm s}(l)e^{-\tau_{\rm 21cm}(R)} \frac{d\tau_{\rm 21cm}(R)}{dR}.
\label{eq:Tb}
\end{equation}
Here $\tau_{\rm 21cm}(R)$ is the 21cm optical depth along the photon path at $R$, which 
is obtained from\footnote{
Here we omitted the optical depth arising from the IGM.
Effects of the optical depth from the IGM in the brightness temperature (Eq.~\eqref{eq:Tb}) are significant only for 
very small minihalo masses, whose contributions in the total brightness temperature in Eqs.~\eqref{eq:dTbbar} and \eqref{eq:rms_dTb}
are irrelevant.
}
\begin{equation}
\tau_{\rm 21cm}(R)=\frac{2c^2A_{10}T_*}{32\pi\nu_0^2}
\int^R_{-\infty} \frac{n_{\rm HI}(l')\phi(\nu',l')}{T_{\rm s}(l')}dR', 
\label{eq:tau_halo}
\end{equation}
where $l'=\sqrt{R^{\prime2}+r^2}$, $A_{10}=2.85\times 10^{-15}$ s$^{-1}$ and $k_BT_*=h\nu_0=5.9\times10^{-6}$ 
eV are respectively the spontaneous decay rate and emitted energy of the 21cm hyperfine transition, and
$n_{\rm HI}$ is the number density of neutral hydrogen atoms. 
The function $\phi(\nu',l')$ is the line profile, which can be modeled by a thermal Doppler-broadening i.e. 
$\phi(\nu',l')=\frac{1}{\sqrt{\pi}\Delta \nu(l')}\exp[-(\frac{\nu'-\nu_0}{\Delta \nu(l')})^2]$ with
$\Delta \nu=(\nu_0/c)\sqrt{2k_BT_K(l')/m_H}$, where $m_H$ is the hydrogen mass.
The optical depth $\tau_{\rm 21cm}$ in Eq.~\eqref{eq:Tb} is given as $\tau_{\rm 21cm}=\tau_{\rm 21cm}(R\to\infty)$.
The differential brightness temperature observed today at frequency
$\nu=\nu'/(1+z)$ can be written as
\begin{equation}
\delta T_b(\nu;z,r,M)=T_b(\nu',z,r,M)/(1+z) -T_{\rm CMB}(0).
\label{eq:dTb}
\end{equation}
The mean surface brightness temperature for a halo with mass $M$ at
$z$ is
provided by
\begin{equation}
\langle \delta T_b \rangle (\nu;z,M)
 = \frac{1}{A(M)} \int dr~ 2\pi r \delta T_b(\nu;z,r,M) 
\end{equation}
where $A(M, z)$ is the geometric cross-section of a halo with mass $M$ at $z$,
$A=\pi r_t^2$.

The differential line-integrated flux from this halo is given by~\cite{Iliev:2002gj}
\begin{equation}
\delta  F = \int d\nu'~ 2 \nu'^2 k_B \langle \delta T_b \rangle(\nu';z,M)
 A(M,z).
\end{equation}
For an optically thin halo, we can replace the integration with the
multiplication by $\Delta\nu_{\rm eff}(z,b)\equiv
\frac{1}{(1+z)\phi(\nu_0,b)}$ at $\nu^\prime=\nu_0$.

Now we consider an observation with a finite bandwidth $\Delta \nu$ and
beam width $\Delta \theta$ .
The mean differential flux per unit frequency is expressed as
\begin{equation}
 \frac{d \delta F}{d \nu} =
\frac{\Delta z(\Delta\Omega)_{\rm beam}}{\Delta\nu}
\frac{d^2V(z)}{dz\,d\Omega}
\int^{M_{\rm max}(z)}_{M_{\rm min}(z)} dM
\delta F
\frac{dn}{dM}(M,z)
\end{equation}
where $(\Delta\Omega)_{\rm beam}=\pi(\Delta\theta/2)^2$,
and $\Delta\nu/\Delta z=\nu_0/(1+z)^2$, and
$M_{\rm max}(z)$ and $M_{\rm min}(z)$
are respectively the maximum and minimum masses contributing to the 21cm line emission.
The maximum mass $M_{\rm max}(z)$ is set to the virial mass $M_*(z)$ in Eq.~\eqref{eq:m*}
for which a virial temperature is $10^4$K.
Above $M_*(z)$, most of the hydrogen in halos is ionized and do not contribute to 21cm line emission
(This is also consistent with the discussion we presented in Section~\ref{sec:opt}). 
On the other hand, the minimum mass is set to the Jeans mass
\begin{equation}
M_{\rm J}(z)=5.73\times 10^3\left(\frac{\Omega_mh^2}{0.15}\right)^{-1}
\left(\frac{\Omega_bh^2}{0.022}\right)^{-3/5}
\left(\frac{1+z}{10}\right)^{3/2}M_\odot.
\end{equation}

Defining the beam-averaged ``effective'' differential antenna temperature $\overline{\delta T}_b$ by
$d \delta F /d \nu=2\nu^2k_B
\overline{\delta T_b}(\Delta\Omega)_{\rm beam}$, we obtain
\begin{equation}
\overline{ \delta T_b}(\nu)\approx
2\pi c\frac{(1+z)^4}{H(z) \nu_0}\int^{M_{\rm max}(z)}_{M_{\rm min}(z)}
dM~\frac{dn}{dM}(M,z) \Delta\nu_{\rm eff}(z)  A(M,z)\langle \delta T_b \rangle(z,M).
\label{eq:dTbbar}
\end{equation}

So far, we have assumed that the halo number density is homogeneous, following
$dn/dM$. However, halos cluster, depending on the density fluctuations,
\begin{equation}
 \delta _N (M) = b(M) \delta ,
\end{equation}
where $\delta_N(M)$ is the number density contrast of halos with mass
$M$ and $b(M)$ is bias.
This clustering causes  fluctuations in $\overline{ \delta T_b}(\nu)$.
Since the rms density fluctuations in a beam are given by
\begin{equation}
\sigma^2(\nu, \Delta \nu, \Delta \theta) =
\int \frac{d^3k}{(2\pi)^3}W(\vec k; \nu,\Delta\nu, \Delta \theta)^2P(k),
\end{equation}
where $W$ is a pencil beam window function at frequency $\nu$ with band width $\Delta \nu$ and beam width $\Delta\theta$
(See e.g. Ref.~\cite{Dodelson:2003ft}), 
the rms halo number density fluctuations with $M$ in a beam is obtained from
\begin{equation}
 \sigma_N(M) = b(M) \sigma(\bar \nu, \Delta \nu, \Delta \theta).
\end{equation}
Therefore, the  rms fluctuations of $\delta T_b(\nu)$ 
are provided by
\begin{equation}
\sigma_{\delta T_b}=\langle |\delta T_b(\nu)|^2 \rangle^{1/2}
\approx \overline{\delta T_b}(\nu)  \beta(z) \sigma(\bar \nu, \Delta \nu, \Delta \theta),
\label{eq:rms_dTb}
\end{equation}
where $\beta(z)$ is the effective bias of the minihalos weighted by their 21cm line fluxes, defined as
\begin{equation}
\beta(z)\equiv \frac{\int^{M_{\rm max}(z)}_{M_{\rm min}(z)}  dM \frac{dn}{dM}(M,z) \mathcal F(z,M) b(M,z)}
{\int^{M_{\rm max}(z)}_{M_{\rm min}(z)}  dM \frac{dn}{dM}(M,z) \mathcal F(z,M)},
\end{equation}
where $\mathcal F(z,M)\equiv\int d^2b~\delta T_b(z,b,M)\propto T_br_t^2 \sigma_V$~\cite{Chongchitnan:2012we} is the flux from a minihalo.

\begin{table}
  \begin{center}
    \begin{tabular}{lr}
      \hline\hline
      antenna collecting area $A$ & $10^5$ m$^2$ \\
      bandwidth $\Delta \nu$ & $1$ MHz \\
      beam width $\Delta \theta$ & $9$ arcmin \\
      integration time $t$ & $10^3$ h \\
      \hline\hline
    \end{tabular}
  \end{center}
  \caption{Survey parameters for the SKA survey}
  \label{tbl:ska}
\end{table}

\begin{figure}
  \begin{center}
  \begin{tabular}{ccc}
    \hspace{-5mm}\scalebox{.4}{\includegraphics{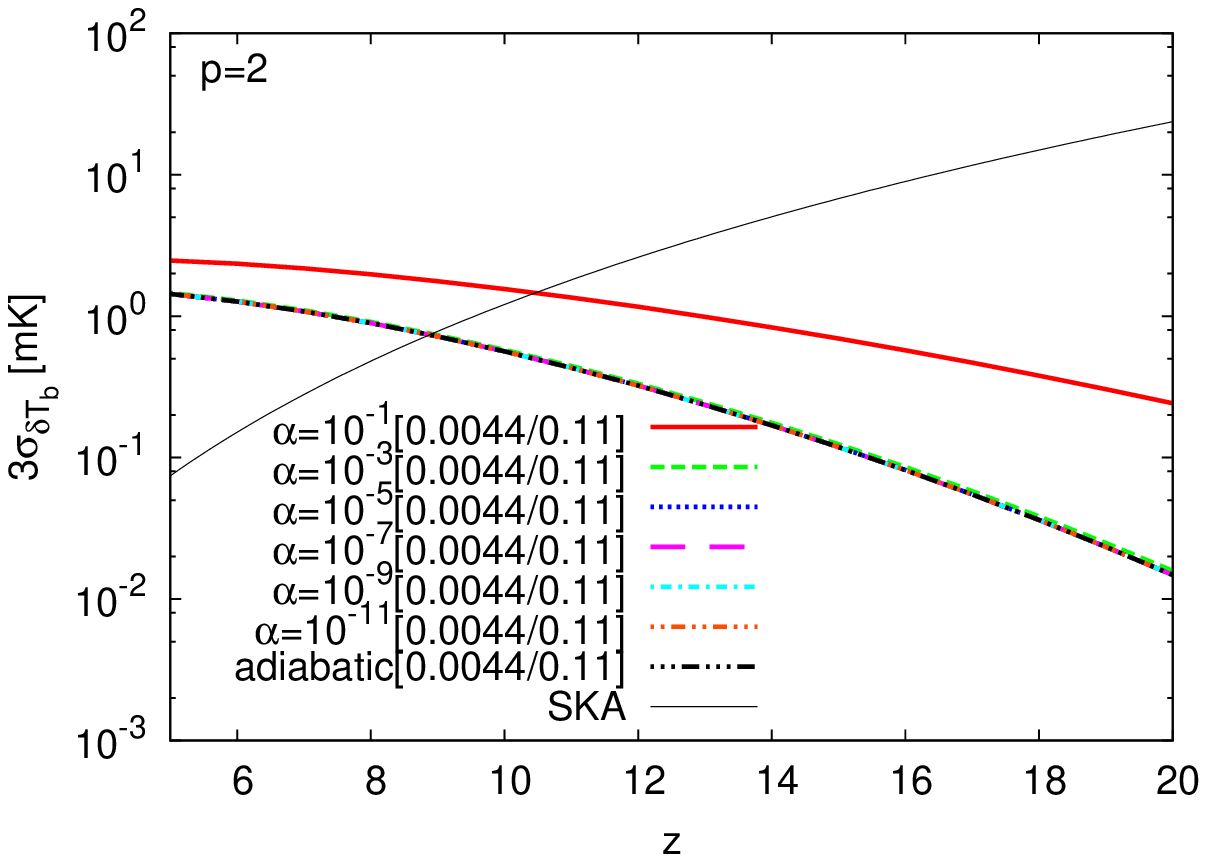}} &
    \hspace{-5mm}\scalebox{.4}{\includegraphics{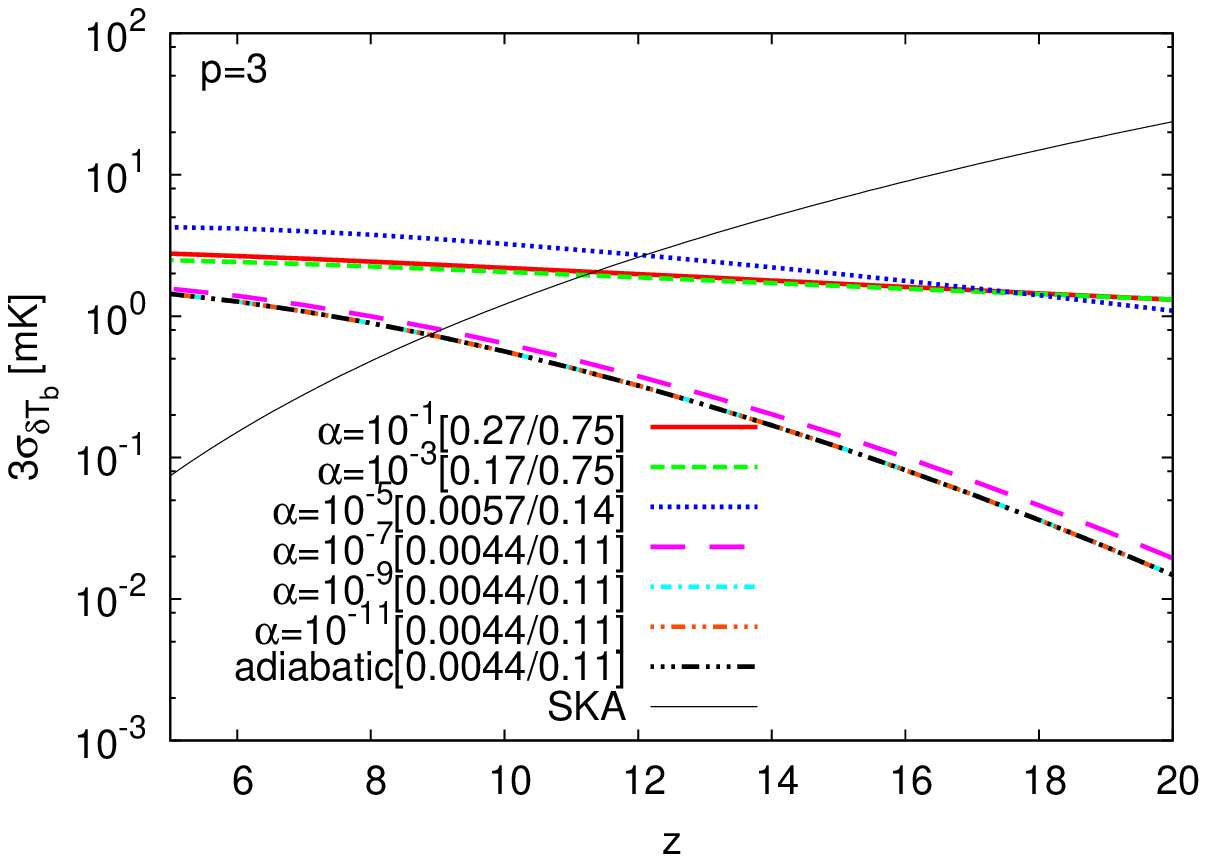}} &
    \hspace{-5mm}\scalebox{.4}{\includegraphics{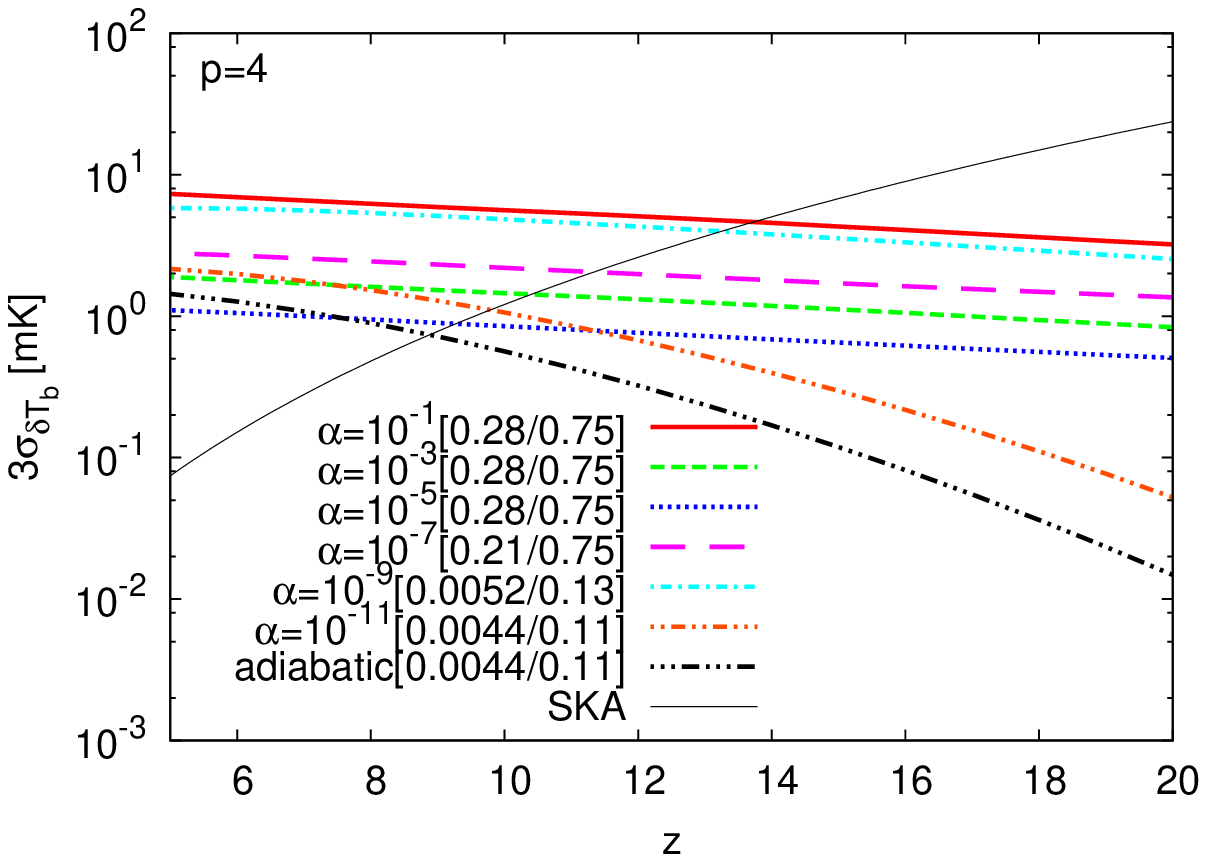}}
  \end{tabular}
  \end{center}
  \caption{Shown are the rms 21cm differential brightness fluctuations $3 \sigma_{\delta T_b}$ for models with 
  blue-tilted isocurvature power spectra as a function of redshift.
  In order from left to right, the spectral index of isocurvature power spectrum $p$ is varied from $2$ to $4$. 
  In each panel, cases of different values of $\alpha=10^{-1}$ (red solid), 
  $10^{-3}$ (green short-dashed), $10^{-5}$ (blue dotted), $10^{-7}$ (magenta long-dashed), 
  $10^{-9}$ (green dot-dashed), $10^{-11}$ (blue two-dot-chain),
  and the pure adiabatic case $\alpha=0$ (black three-dot-chain) are plotted.
  For each parameter set ($p,~\alpha$), we indicate 
  the reionization optical depth computed in Section~\ref{sec:opt} assuming $f_{\rm net}=10^{-6}$ (former) and $10^{-4}$ (latter) 
  in the legend with square bracket.
  As a reference, we also depict the noise level of a SKA-like survey with survey parameters summarized in Table~\ref{tbl:ska} (thin black solid).
  }
  \label{fig:rms}
\end{figure}

In Fig.~\ref{fig:rms}, we plot $3\sigma_{\delta T_b}$
as a function of $6<z=\bar\nu/\nu_0-1<20$ for different parameter values 
for the isocurvature power spectrum $(p,~\alpha)$. 
The adopted resolution and  bandwidth assume a SKA-like survey whose survey parameters are summarized in Table.~\ref{tbl:ska}.
As a reference, the noise level of the SKA-like survey is also depicted in the figure, and is given by~\cite{Furlanetto:2006jb}
\begin{equation}
\sigma_{dT_b}^{\rm (noise)}=20\mbox{mK}\left(\frac{A}{10^4 \mbox{m}^2}\right)^{-1}
\left(\frac{\Delta \theta}{10^\prime}\right)^{-2}
\left(\frac{1+z}{10}\right)^{4.6}
\left(\frac{\Delta \nu}{\mbox{MHz}}\frac{t}{100\mbox{h}}\right)^{1/2},
\end{equation}
where $A$ is the antenna collecting area and $t$ is the integration time.
From the figure, we first see that, for $p=2$, the signal $\sigma_{\delta T_b}$ monotonically increases as $\alpha$ does
(though we cannot distinguish lines for $\alpha<10^{-2}$). On the other hand, for $p=3$ and $p=4$, we can see 
the dependence of $\sigma_{\delta T_b}$ on $\alpha$ is more complicated;
As $\alpha$ increases from zero, $\sigma_{\delta T_b}$ also grows.
When $\alpha$ exceeds some value depending on 
redshift, increment of 
$\alpha$ makes $\sigma_{\delta T_b}$ decrease. However, keeping $\alpha$ increasing, 
one can find that $\sigma_{\delta T_b}$ starts to increase again.
In particular, what may be remarkable is that  
the signal can even be smaller than the vanilla model at low redshifts. 
As we will show in the following, this complicated behavior arises mainly from the mass function which
has a non-trivial dependence on $\alpha$ for large $p\gtrsim3$ as shown in section~\ref{sec:massf}.
We also note that the redshift evolution of $\sigma_{\delta T_b}$ becomes quantitatively different for large values of $\alpha$; 
while $\sigma_{\delta T_b}$ increases as the universe ages
for small $\alpha$, it tends to evolve less for large $\alpha$.

\begin{figure}
  \begin{center}
  \begin{tabular}{ccc}
    \hspace{-5mm}\scalebox{.4}{\includegraphics{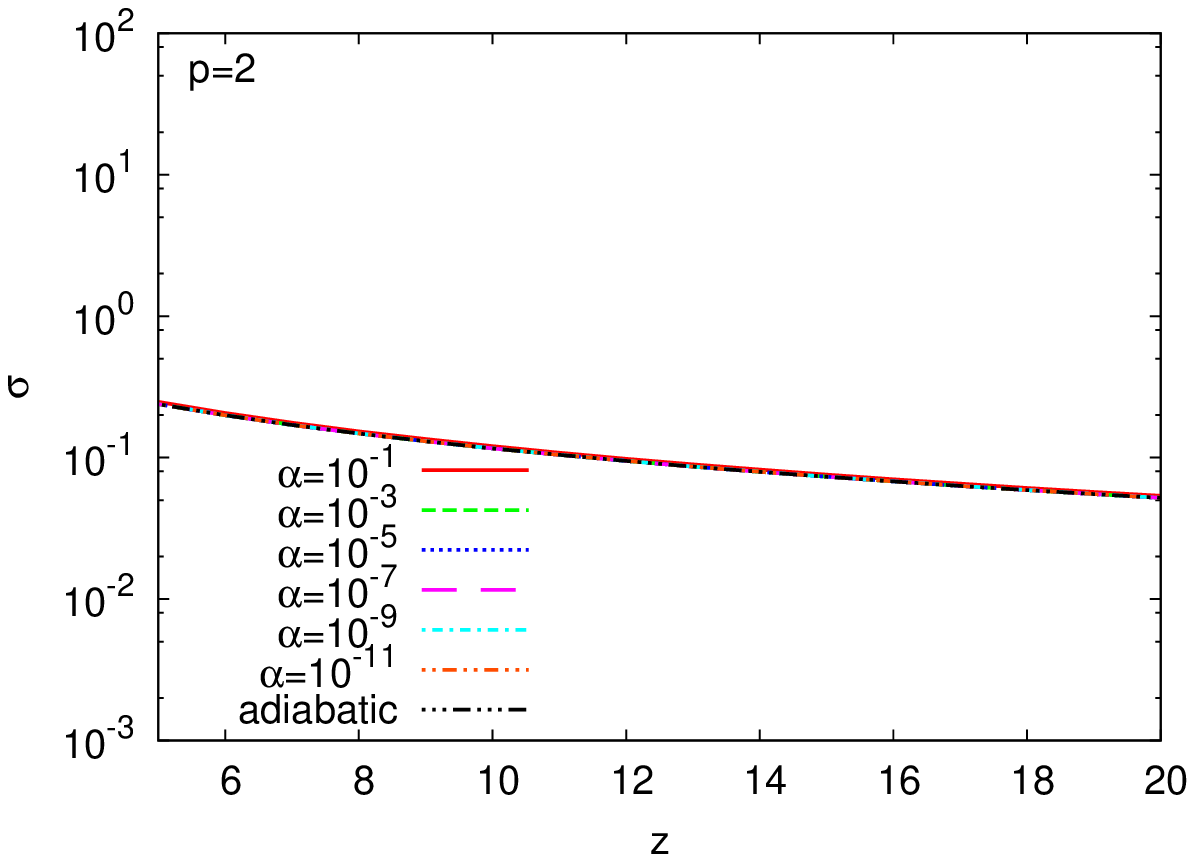}} &
    \hspace{-5mm}\scalebox{.4}{\includegraphics{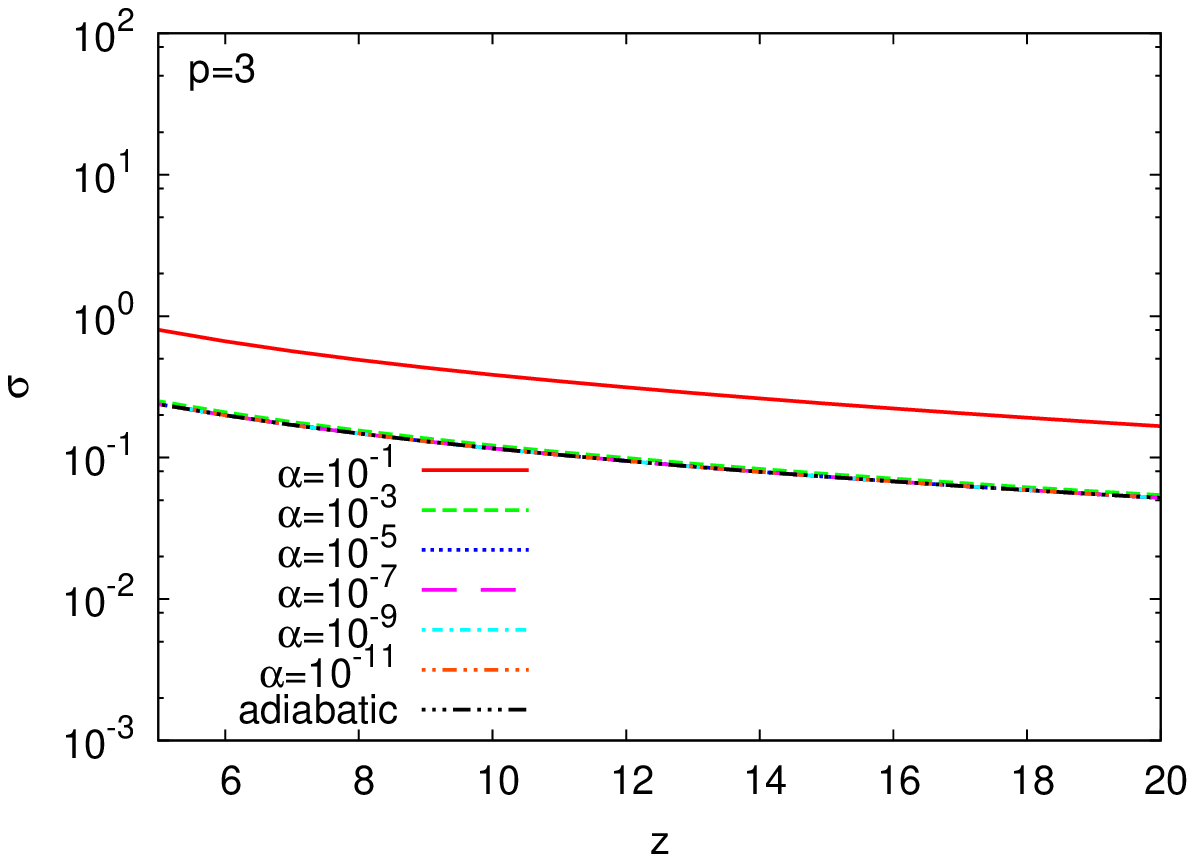}} &
    \hspace{-5mm}\scalebox{.4}{\includegraphics{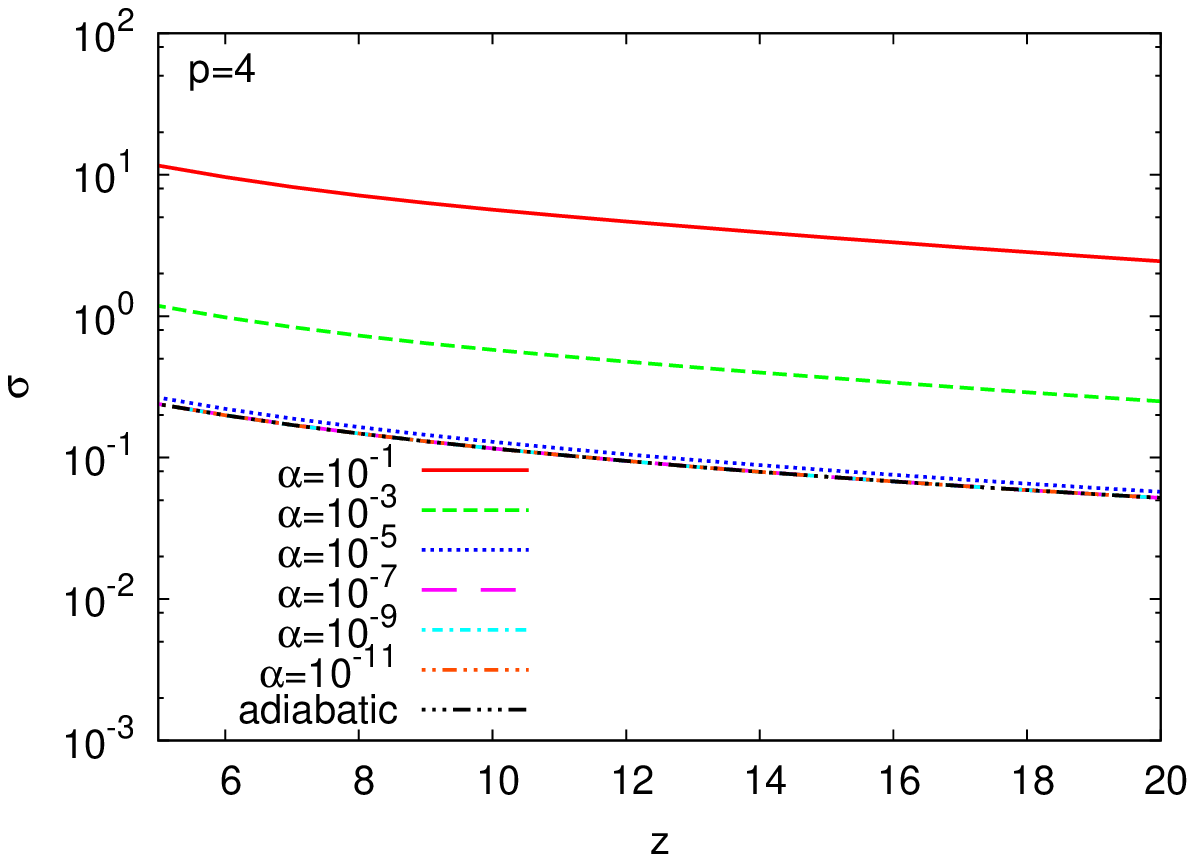}} \\
    \hspace{-5mm}\scalebox{.4}{\includegraphics{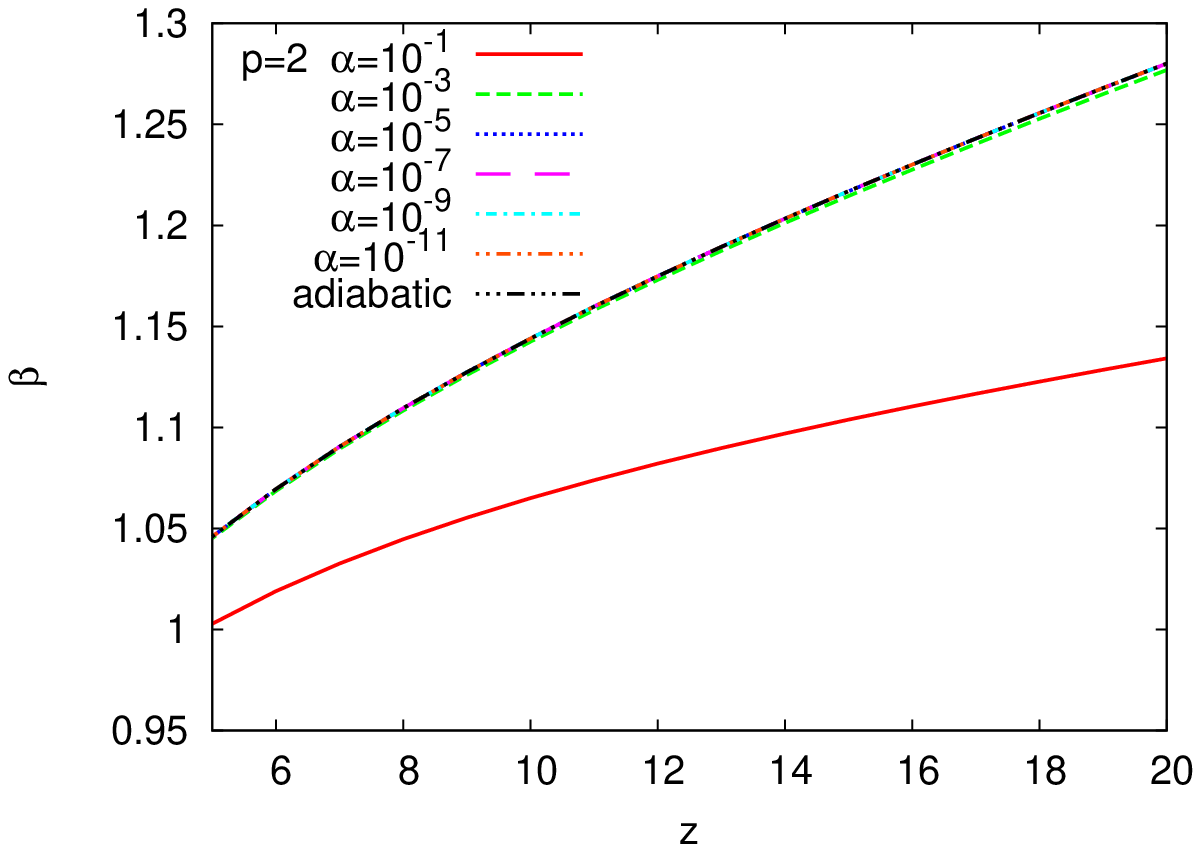}} &
    \hspace{-5mm}\scalebox{.4}{\includegraphics{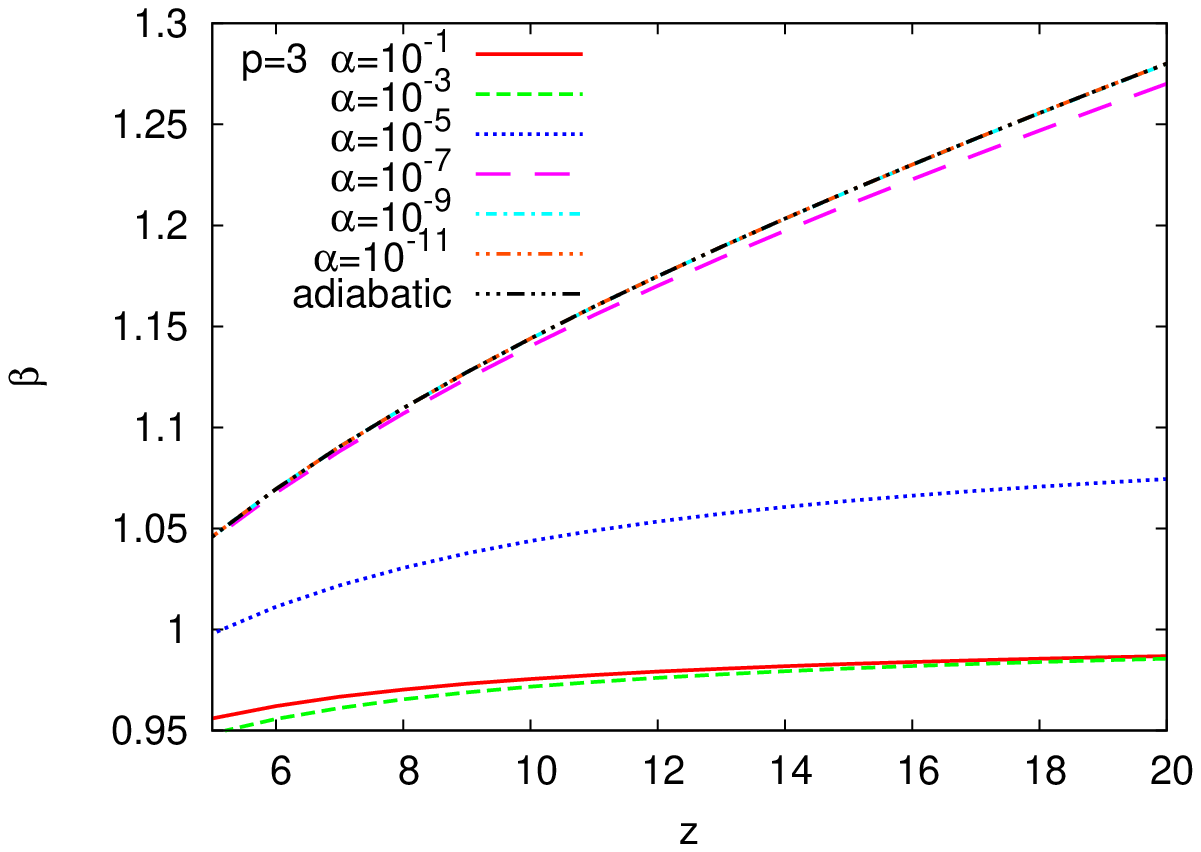}} &
    \hspace{-5mm}\scalebox{.4}{\includegraphics{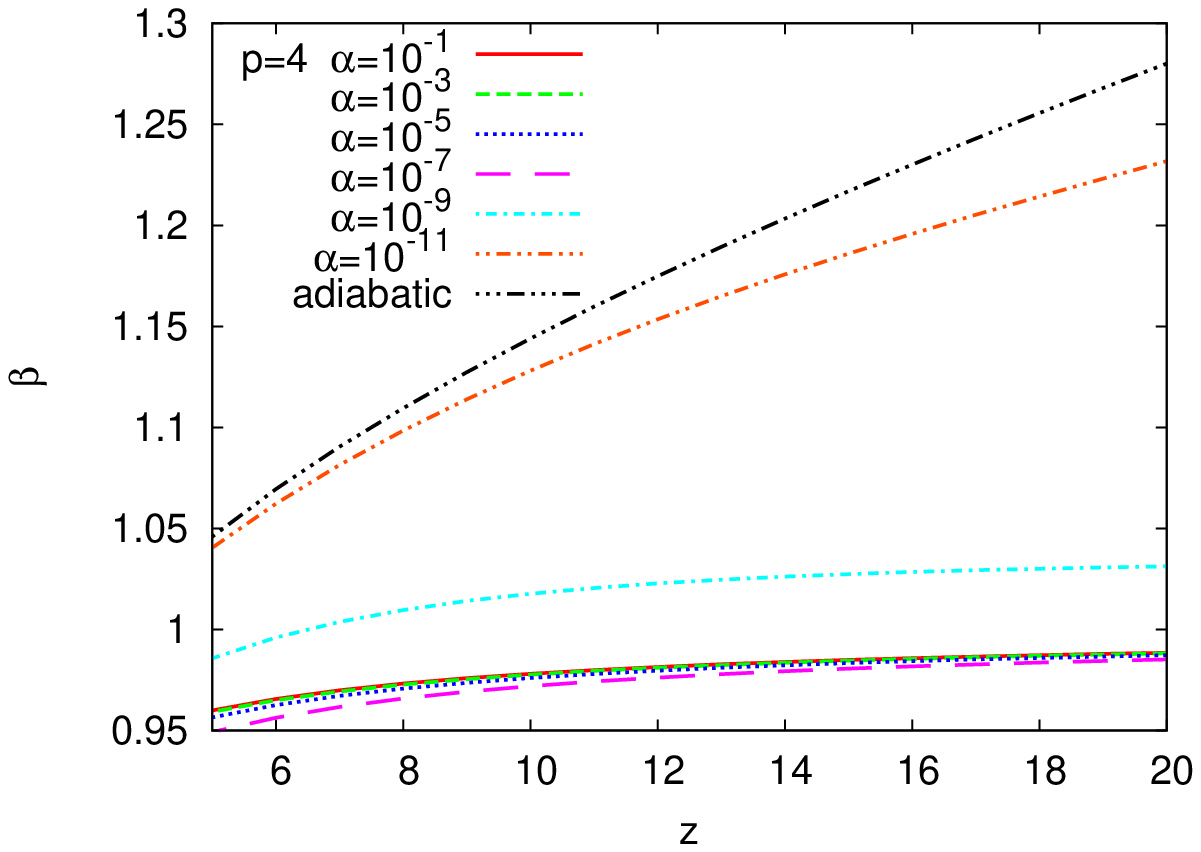}} \\
    \hspace{-5mm}\scalebox{.4}{\includegraphics{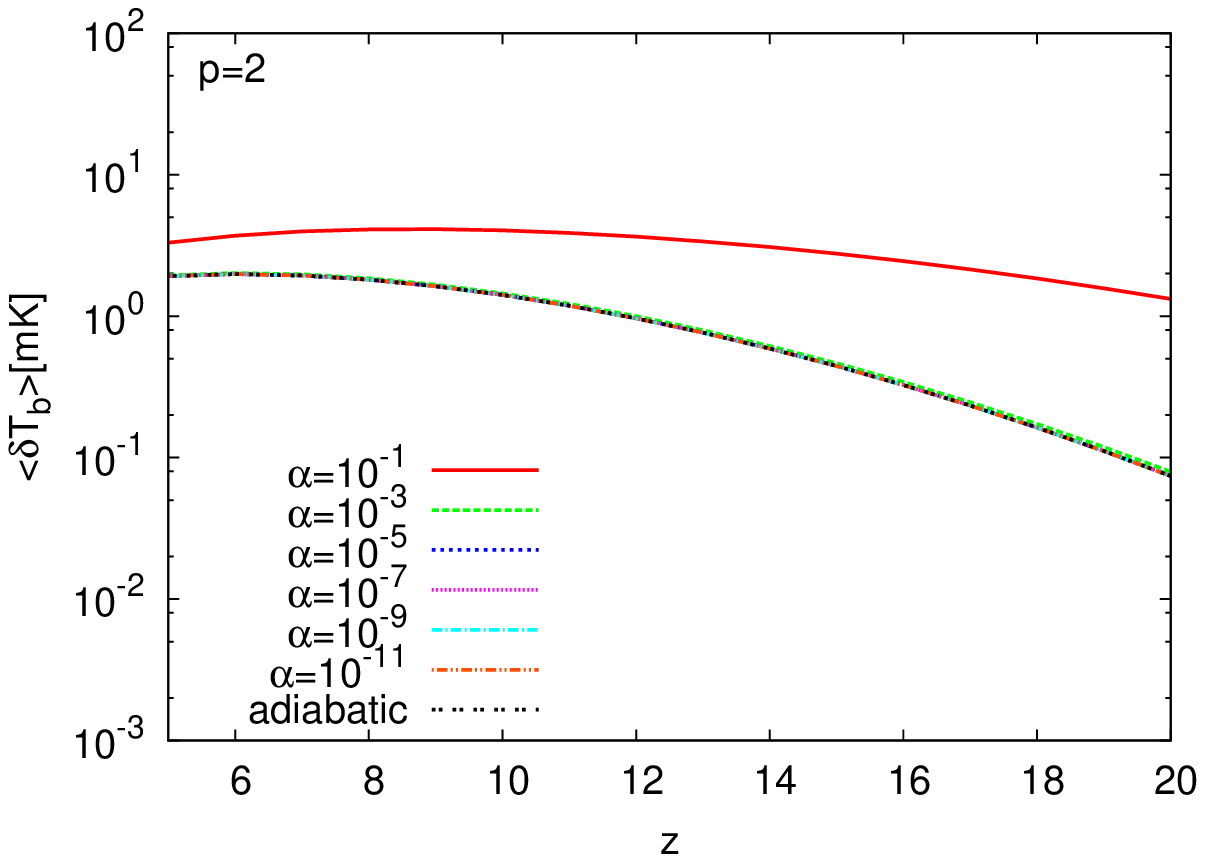}} &
    \hspace{-5mm}\scalebox{.4}{\includegraphics{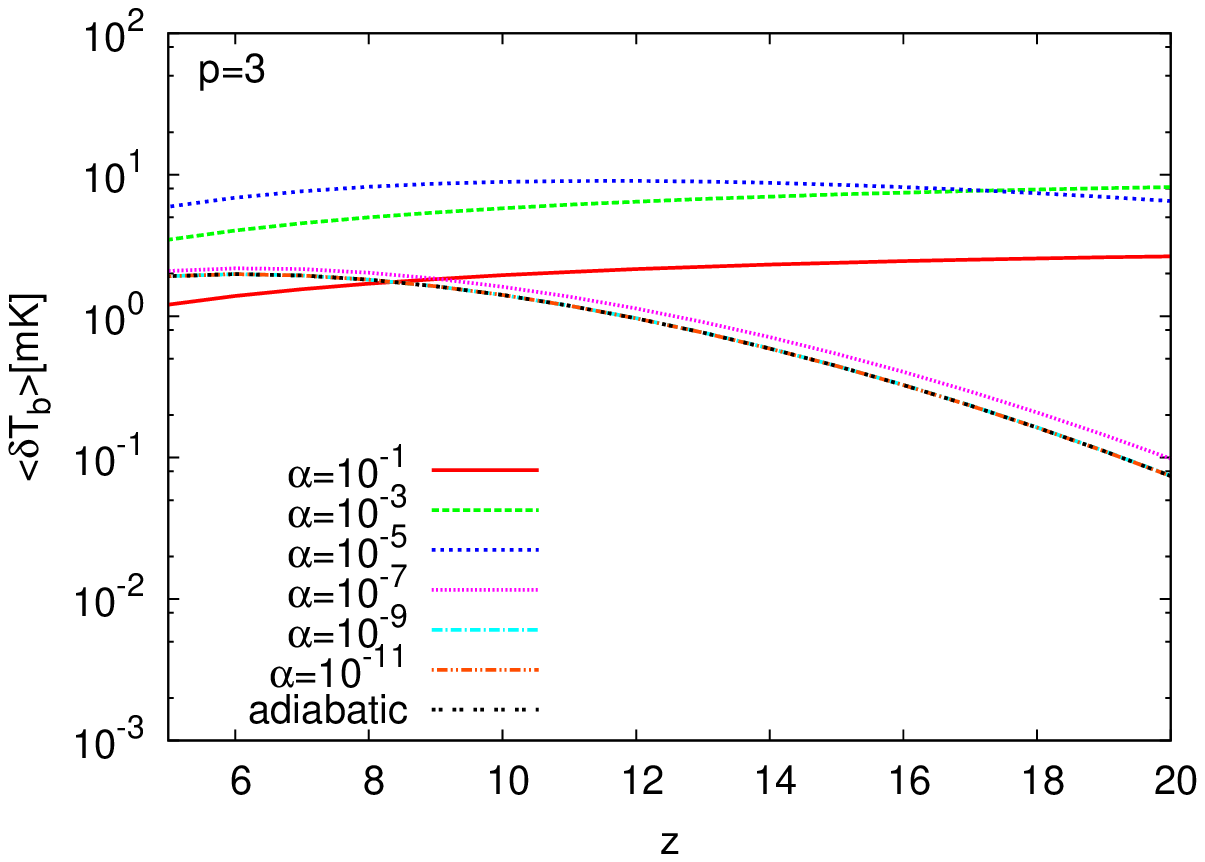}} &
    \hspace{-5mm}\scalebox{.4}{\includegraphics{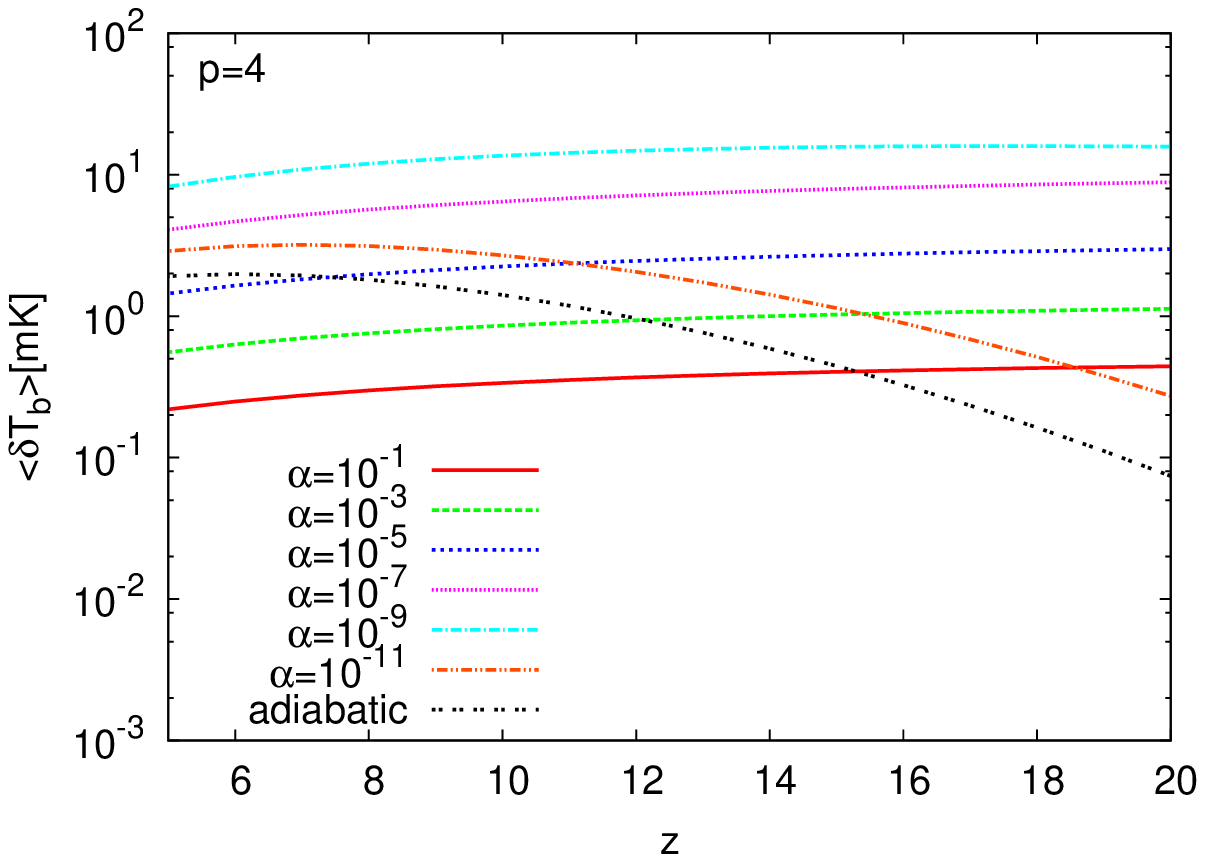}}
  \end{tabular}
  \end{center}
  \caption{The three components in Eq.~\eqref{eq:rms_dTb} are plotted separately. 
  In order from top to bottom, the rms smoothed matter fluctuations $\sigma(\bar\nu,\Delta\nu,\Delta\theta)$,
  the effective bias $\beta(z)$ and the mean value of the 21cm differential brightness temperature
  $\overline{\delta T_b}(\bar \nu)$ are shown as a function of redshift $z$.
  Cases of the spectral index $p=2$, 3 and 4 are shown in order from left to right. 
  In each panel, cases of different values of $\alpha=10^{-1}$ (red solid), 
  $10^{-3}$ (green short-dashed), $10^{-5}$ (blue dotted), $10^{-7}$ (magenta long-dashed), 
  $10^{-9}$ (green dot-dashed), $10^{-11}$ (blue two-dot-chain),
  and the pure adiabatic case $\alpha=0$ (black three-dot-chain) are plotted.
  }
  \label{fig:comp}
\end{figure}

To understand the complicated dependence of $\sigma_{\delta T_b}$ on $\alpha$ and $p$, we divide 
$\sigma_{\delta T_b}$ into parts $\sigma(\nu, \Delta \nu,\Delta \theta)$, $\beta (z)$ and $\overline{\delta T_b}(z)$ 
and investigate each of them separately.
In Fig.~\ref{fig:comp}, we plot the rms of smoothed matter fluctuations $\sigma(\nu, \Delta \nu, \Delta \theta)$, 
the effective bias $\beta(z)$ and the mean 21cm brightness temperature $\overline{\delta T_b}(\nu)$ 
as a function of $z=\nu_0/\nu-1$ for different values of $p$ and $\alpha$.
Here $\Delta \nu$ and $\Delta \theta$ are fixed to the values adopted in Fig.~\ref{fig:rms}.

As expected, $\sigma(\nu, \Delta \nu, \Delta \theta)$ is monotonically enhanced as $\alpha$ increases. 
If we fix $\alpha$, the enhancement is more significant for larger $p$. 
However, as long as $\alpha$ is not so large as to be excluded by current observations (See Figs.~\ref{fig:loglike}), 
the enhancement is not significant and at most $\mathcal O(1)$.
We can also see that $\beta(z)$ is not significantly affected by the isocurvature power spectrum; 
for $2<p<4$, $\beta(z)$ changes at most 30 \% as long as $\alpha<0.1$.

On the other hand, as seen in the figure, $\overline{\delta T_b}(\nu)$ has a complex response to $\alpha$ depending on $p$.
For a moderately blue-tilted isocurvature spectrum with $p=2$, $\overline{\delta T_b}(\nu)$ monotonically increases as $\alpha$ increases.
On the other hand, for $p=3$ and $p=4$, the dependence of $\overline{\delta T_b}(\nu)$ on $\alpha$ dramatically changes.
As $\alpha$ increases from zero, while at first $\overline{\delta T_b}(\nu)$ increases, then it begins to decrease at some value of $\alpha$, 
which depends on $p$. Since the isocurvature power spectrum affects $\overline{\delta T_b}(\nu)$ only through the mass function, 
this behavior can be understood by looking at the mass function $dn/d\ln M$ with $M_{\rm J}(z)<M<M_*(z)$.
As we discussed in Section~\ref{sec:massf}, an
increase in $\alpha$ does not necessarily increase $dn/d\ln M$ for a highly blue-tilted isocurvature power spectrum
and, in particular, $dn/\ln M$ for large $\alpha$ can be below that for the vanilla model $\alpha=0$ 
at relatively small masses $10^3M_\odot \lesssim M\lesssim10^7M_\odot$ (this roughly corresponds to $M_{\rm J}(z)<M<M_*(z)$
for $z\simeq10$). Since minihalos which contribute to the 21cm brightness temperature are limited to those with these small masses, 
$\overline{\delta T_b}(\nu)$ reflects this complicated behavior of the mass function.

To summarize, the complicated dependence of $\sigma_{\delta T_b}$ on $\alpha$ is understood as follows.
A blue-tilted isocurvature power spectrum affects $\sigma_{\delta T_b}$ mainly through the rms smoothed mass 
fluctuations $\sigma(\nu,\Delta \nu,\Delta \theta)$ and the mean differential brightness temperature $\overline{\delta T_b}(\nu)$.
For a moderately blue-tilted isocurvature spectrum $p=2$, $\sigma(\nu,\Delta \nu,\Delta \theta)$ and $\overline{\delta T_b}$
increase monotonically as $\alpha$ increases.
On the other hand, for a highly blue-tilted spectrum $p=3$ or $p=4$, while 
$\sigma(\nu,\Delta \nu,\Delta \theta)$ continuously monotonically increases  as $\alpha$ increases, 
$\overline{\delta T_b}$ decreases at some $\alpha$. 
As long as $\alpha$ is not so large as to be excluded by current observations, 
$\sigma_{\delta T_b}$ first increases and then  decreases, mainly reflecting the dependence of $\overline{\delta T_b}$.
When $\alpha$ is significantly large so that it is already excluded by current observations, the 
influence of $\sigma(\nu,\Delta \nu, \Delta \theta)$ dominates and $\sigma_{\delta T_b}$ again starts to increase.


\begin{figure}
  \begin{center}
    \hspace{-5mm}\scalebox{1.}{\includegraphics{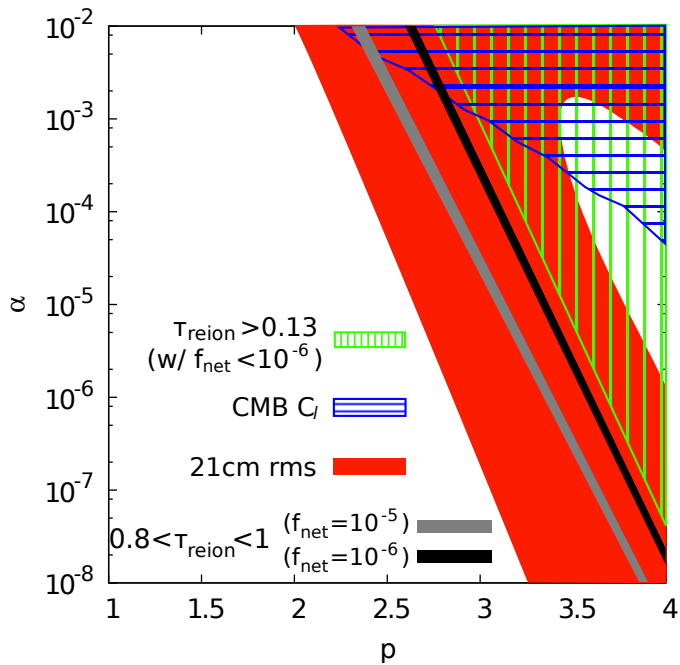}}
  \end{center}
  \caption{Forecast of constraints on the isocurvature power spectrum from observations of 21cm line fluctuations.
  The 2$\sigma$ constraint expected from SKA is shown as red shaded region.
  The 2$\sigma$ constraints from WMAP9+ACT2008 (region shaded with blue horizontal stripes) 
  and reionization optical depth $\tau_{\rm reion}>0.13$ for $f_{\rm net}=10^{-6}$ (region shaded with green vertical stripes) 
  are also shown. 
  Black and gray bands show the expected Planck measurements $\tau_{\rm reion}=0.09\pm0.01$ 
  (2$\sigma$) for $f_{\rm net}=10^{^5}$ and $10^{-6}$.
  }
  \label{fig:s2n}
\end{figure}

Finally, we compute $\Delta \chi^2$, assuming that SKA can observe 21cm line fluctuations over $5\le z\le20$.
Here we simply assume that each band in the observed frequency range is 
independent and has the same survey parameters as in Table~\ref{tbl:ska}.
Then $\Delta \chi^2$ can be given as
\begin{equation}
\Delta \chi^2(p,\alpha)=\frac12
\sum_{\rm bands} \left[\frac{{\sigma_{\delta T_b}}^2(p,\alpha)-{\sigma^{\rm (fid)}_{\delta T_b}}^2}
{{\sigma^{\rm (fid)}_{\delta T_b}}^2+{\sigma^{\rm (noise)}_{\delta T_b}}^2}\right]^2, 
\label{eq:s2n}
\end{equation}
where $\sigma^{\rm (fid)}_{\delta T_b}$ denotes the rms 21cm line emission fluctuations in the case of the fiducial pure adiabatic model.
In Fig.~\ref{fig:s2n}, we plot 
the 2$\sigma$ constraint expected from the SKA survey.
As a reference, the 2$\sigma$ constraints from the reionization optical depth with a conservative efficiency parameter $f_{\rm net}=10^{-6}$ 
and the CMB power spectrum are also shown here.
In the figure, one can see that the 2$\sigma$ excluded region from 21cm line fluctuations has a complex geometry
and a region with relatively large $\alpha\sim 10^{-6}$-$10^{-3}$ with $p\gtrsim 3.5$ is not excluded by 21cm line observations alone.
This reflects the complicated dependence of the 21cm line fluctuations on the isocurvature power spectrum. 
However, the constraint from the reionization optical depth covers the region. 
This shows that the 21cm line fluctuations and the reionization optical depth can be complementary as probes of a 
blue-tilted isocurvature power spectrum. 
Combining the constraint from the WMAP determination of $\tau_{\rm reion}$ with $f_{\rm net}>10^{-6}$, 
we found that the SKA is expected to give a constraint
\begin{equation}
\log_{10}\alpha<-4.8p+7.7, 
\end{equation}
at the 2$\sigma$ level.
In the same figure, we also plot bands of regions corresponding to $\tau_{\rm reion}=0.09\pm0.01$ (2$\sigma$), 
which is expected to be achieved by Planck observations of the CMB polarization spectrum~\cite{Planck:2006aa}\footnote{
SKA can also constrain $\tau_{\rm reion}$ from the 21cm fluctuations from neutral hydrogen in IGM.
However, such fluctuations are expected to be orders of magnitude smaller than those from minihalos
and we expect the Planck estimate of $\tau_{\rm reion}$ would not be much improved.
This is consistent with the study of Ref.~\cite{Mao:2008ug}.}
for the values of $f_{\rm net}=10^{-5}$ and $10^{-6}$, which correspond to the cases where 
isocurvature perturbations at small scales are responsible for inducing cosmological reionization. 
From the figure, one can see that in these bands, 21cm line fluctuations can be detected with significance of more than 2$\sigma$, 
and thus by observing 21cm line emission, we will have a chance to test whether the isocurvature perturbations are responsible for reionization.

\section{Conclusions}
\label{sec:discussion}
In this paper, we have studied cosmological signatures of a blue-tilted isocurvature power spectrum
with spectral index $2\lesssim p\lesssim4$.
Such an isocurvature power spectrum modifies the mass function of dark matter halos.
This modification in mass function consequently affects both the reionization history and 21cm line fluctuations
from neutral hydrogen in minihalos at high redshifts. 
The presence of a blue-tilted isocurvature spectrum enhances the number density of massive halos at high redshifts
and hence promotes formation of galaxies and reionization of the universe. Given the reionization optical depth estimated 
from current CMB observations, we can obtain constraints on the amplitude and spectral index of the isocurvature power spectrum.
Although the constraints strongly depend on the efficiency of reionization, for reasonable parameter values it can surpass
those obtained from the spectral shape of the CMB power spectrum. On the other hand, a blue-tilted
isocurvature power spectrum changes the mass function at smaller masses in a complicated manner, 
which results in either an enhancement or even a suppression of 21cm line fluctuations depending on the
parameters of  the isocurvature power spectrum. Future surveys such as SKA
can probe the parameter space of isocurvature power spectra which cannot be explored directly from 
the CMB power spectrum. We emphasize that the reionization optical depth and 21cm line fluctuations can provide 
complementary probes of blue-tilted isocurvature power spectra. 

\bigskip
Note added: 
While we were finishing the present work, we noticed that Ref. \cite{Takeuchi:2013hza}, 
which appears on arXiv around the similar time as ours, has some overlap with our 
analysis on 21cm line fluctuations from a blue-tilted power spectrum. 

\acknowledgments
We would like to thank Yoshitaka Takeuchi and Sirichai Chongchitnan for helpful discussion. HT is supported by the DOE at ASU. 
TS is supported by Japan Society for the promotion of Science and the Academy of Finland grant 1263714.
NS is supported by the Grant-in-Aid for the Scientific Research Fund under Grant No.~25287057.
This work is also supported by the World Premier International Research Center Initiative (WPI Initiative), MEXT, Japan.
The research of JS has been supported at IAP by  the ERC project  267117 (DARK) 
hosted by Universit\'e Pierre et Marie Curie - Paris 6   and at JHU by NSF grant OIA-1124403.

\providecommand{\href}[2]{#2}\begingroup\raggedright

\end{document}